\documentclass[aps,showpacs,twocolumn,superscriptaddress,floatfix,prc]{revtex4-1}
\usepackage{dcolumn}   
\usepackage{graphicx}
\usepackage{amsfonts}
\usepackage{amsmath,amssymb}
\usepackage{bm}
\usepackage{subfigure}
\usepackage{color} 

\newcommand{\beqn}{\begin{equation}}
\newcommand{\eeqn}{\end{equation}}
\newcommand{\beq}{\begin{equation}}
\newcommand{\eeq}{\end{equation}}
\newcommand{\bea}{\begin{eqnarray}}
\newcommand{\eea}{\end{eqnarray}}
\newcommand{\ba}{\begin{align}}
\newcommand{\ea}{\end{align}}

\newcommand{\rvec}{{\bf r}}

\newcommand{\Rvec}{{\bf R}}

\newcommand{\mpi}{m_{\pi}}

\begin{document}
\title{Microscopically-based energy density functionals for nuclei \\
using the density matrix expansion: \\Implementation and pre-optimization}

\author{M.~Stoitsov}
\email{stoitsovmv@ornl.gov}
\affiliation{Department of Physics \& Astronomy, University of Tennessee, Knoxville, Tennessee 37996, USA \\
Physics Division,  Oak Ridge National Laboratory, Oak Ridge, Tennessee 37831, USA}

\author{M.~Kortelainen}
\email{kortelainene@ornl.gov}
\affiliation{Department of Physics \& Astronomy, University of Tennessee, Knoxville, Tennessee 37996, USA \\
Physics Division,  Oak Ridge National Laboratory, Oak Ridge, Tennessee 37831, USA}


\author{S.~K.~Bogner}
\email{bogner@nscl.msu.edu}
\affiliation{National Superconducting Cyclotron Laboratory,
             1 Cyclotron Laboratory,
             East-Lansing, MI 48824, USA}
\affiliation{Department of Physics and Astronomy,
             Michigan State University, East Lansing, MI 48824, USA}

\author{T.~Duguet}
\email{thomas.duguet@cea.fr}
\affiliation{National Superconducting Cyclotron Laboratory,
             1 Cyclotron Laboratory,
             East-Lansing, MI 48824, USA}
\affiliation{Department of Physics and Astronomy,
             Michigan State University, East Lansing, MI 48824, USA}
\affiliation{CEA, Centre de Saclay, IRFU/Service de Physique Nucl´eaire, F-91191 Gif-sur-Yvette, France}

\author{R.~J.~Furnstahl}
\email{furnstahl.1@osu.edu}
\affiliation{Department of Physics, Ohio State University, Columbus, OH 43210, USA}

\author{B.~Gebremariam}
\email{gebremar@nscl.msu.edu}
\affiliation{National Superconducting Cyclotron Laboratory,
             1 Cyclotron Laboratory,
             East-Lansing, MI 48824, USA}
\affiliation{Department of Physics and Astronomy,
             Michigan State University, East Lansing, MI 48824, USA}

\author{N.~Schunck}
\email{schuncknf@ornl.gov}
\affiliation{Department of Physics \& Astronomy, University of Tennessee, Knoxville, Tennessee 37996, USA \\
Physics Division,  Oak Ridge National Laboratory, Oak Ridge, Tennessee 37831, USA}

\date{\today}

\begin{abstract}
In a recent series of papers, Gebremariam, Bogner, and Duguet derived a microscopically based nuclear energy density functional by applying the Density Matrix Expansion (DME) to the Hartree-Fock energy obtained from chiral effective field theory (EFT) two- and three-nucleon interactions. Due to the structure of the chiral interactions, each coupling in the DME functional is given as the sum of a coupling constant arising from zero-range contact interactions and a coupling function of the density arising from the finite-range pion exchanges. Since the contact contributions have essentially the same structure as those entering  empirical Skyrme functionals, a microscopically guided Skyrme phenomenology has been suggested in which the contact terms in the DME functional are released for optimization to finite-density observables to capture short-range correlation energy contributions from beyond Hartree-Fock. The present paper is the first attempt to assess the ability of the newly suggested DME functional, which has a much richer set of density dependencies than traditional Skyrme functionals, to generate sensible and stable results for nuclear applications. The results of the first proof-of-principle calculations are given, and numerous practical issues related to the implementation of the new functional in existing Skyrme codes are discussed.  Using a restricted singular value decomposition (SVD) optimization procedure, it is found that the new DME functional gives numerically stable results and exhibits a small but systematic reduction of our test $\chi^2$ function compared to standard Skyrme functionals, thus justifying its suitability for future global optimizations and large-scale calculations.
\end{abstract}


\maketitle

\section{Introduction}
\label{sec0}

One of the fundamental challenges of nuclear theory is to predict properties of nuclei starting from the underlying vacuum two- and three-nucleon interactions. While impressive progress has been made in extending the limits of ab-initio methods beyond the lightest nuclei~\cite{Pieper:2004qh,Quaglioni:2007qe,Hagen:2008iw}, the nuclear energy density functional (EDF) approach is the only computationally feasible many-body method capable of describing nuclei across the mass table~\cite{bender03b}. Driven by interest in the coming generation of radioactive isotope beam facilities, along with studies of astrophysical systems such as neutron stars and supernovae that require controlled extrapolations of nuclear properties in isospin, density, and temperature, there is a large effort currently under way to develop nuclear energy functionals with substantially reduced global errors and improved predictive power away from stability. The Universal Nuclear Energy Density Functional  (UNEDF) SciDAC-2 collaboration is one such effort that aims to develop a comprehensive theory of nuclear structure and reactions utilizing the most advanced computational resources and algorithms available, including high-performance computing techniques to scale to petaflop platforms and beyond \cite{unedf:2007}.

Well-known empirical Skyrme and Gogny EDFs are typically characterized by 10--15 coupling constants adjusted to selected experimental data. Despite their simplicity, such functionals provide a remarkably good description of a broad range of bulk properties such as ground-state masses, separation energies, etc., and to a lesser extent of certain spectroscopic features of known nuclei. They are also widely employed, with some sucess, in studies of complex nuclear phenomena such as, e.g., large-amplitude collective motion. However, their phenomenological nature often leads to parameterization-dependent predictions and does not offer a clear path toward systematic improvements.

One possible strategy is to provide microscopic constraints on the analytical form of the functional and the values of its couplings from many-body perturbation theory (MBPT) starting from the underlying two- (NN) and three-nucleon (NNN) interactions~\cite{Lesinski:2008cd,Drut:2009ce,Duguet:2009gc,Bogner:2008kj,Kaiser:2003uh,Kaiser:2009me,Kaiser:2010pp}. Recent progress in evolving chiral effective field theory (EFT) interactions to lower momentum using renormalization group (RG) methods~\cite{Bogner:2005sn,Bogner:2006vp,Bogner:2006pc,Bogner:2009un,Bogner:2009bt} (see also \cite{Roth:2005ah,Roth:2008km}) is expected to play a significant role in this effort, as the Hartree-Fock approximation becomes a reasonable (if not quantitative) starting point. This suggests that the theoretical developments and phenomenological successes of EDF methods for Coulomb systems may be applicable to the nuclear case for low-momentum interactions.

However, even with these simplifications, the MBPT energy expressions are written in terms of density matrices and propagators folded with finite-range interaction vertices, and are therefore non-local in both space and time. In order to make such functionals numerically tractable in heavy open-shell nuclei, it is necessary to develop simplified approximations, for example based on the use of local densities and currents. At lowest order in MBPT (i.e., Hartree-Fock), the density matrix expansion (DME) of Negele and Vautherin~\cite{negele72} provides a convenient framework to approximate the spatially non-local Fock energy as a local Skyrme-like functional with density-dependent couplings. This novel density dependence of the couplings is a consequence of the finite-range of the vacuum interactions, and is controlled by the longest-ranged components. Consequently, the DME can be used to map physics associated with long-range one- and two-pion exchange interactions into a local EDF form that can be implemented at minimal cost in existing Skyrme codes. The rich spin and isospin structure of such interactions should improve the quantitative predictive power of EDFs while in the same time retaining the connection of these functionals with the underlying microscopic theory of nuclear forces.

We do not expect dramatic changes for bulk nuclear properties due to the tendency of pions to average out in spin and isospin sums, but we do expect interesting consequences for single-particle properties (which phenomenology tells us are sensitive probes of the tensor force) and systematics along long isotopic chains (which should be sensitive to the isovector physics coming from pion-exchange interactions). Another potentially significant advantage of the DME functional is that two very different microscopic origins of spin-orbit properties (i.e., short-range NN and long-range NNN spin-orbit interactions) are treated on an equal footing. This is in contrast to empirical Skyrme and Gogny functionals, where the density-independent spin-orbit coupling is consistent with the short-range NN spin-orbit interaction, but has no obvious connection with the sub-leading (but quantitatively significant) long-range NNN sources of spin-orbit physics. Although it is beyond the scope of the present paper, let us mention that a clear priority of future studies will therefore be to examine if the DME-based functional is able to improve two major shortcomings of standard Skyrme phenomenology~\cite{Lesinski:2007zz}: (i) the destructive interplay between tensor and spin-orbit terms that compromise spin-orbit splittings and the evolution of nuclear shell with isospin, and (ii) the too-high location of high-$l$ centroids compared to low-$l$ ones, which compromise the shell position even for nuclei near the stability valley. Any positive change regarding these two points would significantly impact the performance and predictive power of EDF calculations dedicated to spectroscopy.

Recently, Gebremariam {\it et al.} have used an improved formulation of the DME~\cite{Gebremariam:2009ff} to construct a non-empirical Hartree-Fock energy functional from un-evolved chiral EFT two- and three-nucleon interactions through next-to-next-to-leading-order (N$^2$LO)~\cite{Gebremariam:2010ni,Gebremariam:2009xp}. The structure of the EFT interactions implies that each coupling in the DME HF functional can be written as the sum of a density-independent (Skyrme-like) coupling constant arising from the zero-range contact interactions and a density-dependent coupling function arising from the long-range pion-exchange interactions. As discussed in Section~\ref{sec1}, in the present approach the separation of long- and short-distance physics at the HF level is used to motivate a semi-phenomenological functional.
In particular, the DME coupling functions arising from finite-range pion-exchanges are not modified, while the density-independent couplings associated with the contact interactions are released for optimization to infinite nuclear matter and finite nuclei properties in order to mimic higher-order short-range correlation energy contributions.

It is expected that the semi-phenomenological DME-based functional should perform {\it at least} as well as empirical Skyrme functionals because one still fits the same Skyrme coupling constants to data, the only difference being that the new EDF contains additional parameter-free coupling functions derived from the finite-range NN and NNN interactions.  However, due to the highly non-trivial density-dependence carried by the DME couplings, there is no {\it a priori} guarantee that the implementation will not be plagued with numerical instabilities or other technical difficulties that invalidate the approach.  Consequently, the main goal of the present paper is to perform ``proof-of-principle'' calculations in which we (i) give the practitioner's view of how the DME functional can be implemented in existing Skyrme codes, and (ii) perform a restricted SVD optimization (``pre-optimization'') of the density-independent couplings to verify that the new microscopically guided phenomenology does no worse than standard Skyrme functionals, thus justifying its suitability for future global optimizations and large-scale EDF calculations.

In our initial investigation, we constrain zero-range volume parameters of the DME functional to reasonable values for equilibrium characteristics of infinite nuclear matter (INM), while zero-range surface parameters are obtained from a restricted optimization procedure using SVD techniques based on 72 even-even nuclei binding energies,  and 8 odd-even mass  (OEM) differences (4 neutron and 4 proton).  Our analysis will show that while the DME-based functional is indeed more susceptible to instabilities, it can still be made sufficiently stable to carry out an optimization procedure for nuclei throughout the nuclear mass chart. It should be stressed that a detailed comparison of the quality of the DME functional against standard Skyrme predictions, and {\it a fortiori} experimental data, is premature before applying a more rigorous global optimization. 

The rest of the paper is organized as follows. In Section~\ref{sec1}, we review how the DME can be used to map ab-initio MBPT energy expressions into the form of a local EDF and motivate the semi-phenomenological approach used in the present work. The explicit form of the functional is given in Section~\ref{sec2}, and the free parameters entering the volume part of the functional are expressed in terms of infinite nuclear matter (INM) equilibrium characteristics in Section~\ref{sec3}.  The restricted SVD optimization procedure used to fix free surface parameters in the functional is described in Section~\ref{sec6}, and a comparison of selected nuclear properties calculated with both the DME functional and the standard Skyrme functional is made in Section~\ref{sec7}. Conclusions are given in Section~\ref{sec8}, while various formulas and technical details are collected in the Appendices.

\section{Microscopically motivated functional} \label{sec1}

\subsection{DME exchange energy functional}

At lowest order, the Fock energy computed from un-evolved chiral interactions exhibits spatial non-localities due to the convolution of finite-range form factors with non-local density matrices. The central idea of the DME is to factorize the non-locality of the one-body density matrix (OBDM) by expanding it into a finite sum of terms that are separable in relative $\bm{r}\equiv \bm{r}_1-\bm{r}_2$ and center of mass $\bm{R}\equiv (\bm{r}_1+\bm{r}_2)/2$ coordinates. Adopting notations similar to those introduced in
Refs.~\cite{Gebremariam:2009ff,Gebremariam:2010ni}, one expands the spin-scalar and spin-vector parts (in both isospin channels) of the density matrix as
\begin{eqnarray}
\rho_{t} (\bm{r}_1, \bm{r}_2)  &\approx&  \sum^{n_{\text{max}}}_{n=0} \Pi_n (k  r)
   \,\, {\cal P}_n (\bm{R}) \;, \label{approxscalar}\\
\bm{s}_t (\bm{r}_1, \bm{r}_2) &\approx& \sum^{m_{\text{max}}}_{m=0} \Pi_m (k r)
   \,\, {\cal Q}_m (\bm{R}) \;,\label{approxvector}
\end{eqnarray}
where $k$ is an arbitrary momentum that sets the scale for the decay in the off-diagonal direction, whereas $\Pi_n (k r)$ denotes the so-called
$\Pi-$functions that depend on the particular formulation of the DME, see Refs.~\cite{Gebremariam:2009ff,Gebremariam:2010ni}. In the present work,  $k$ is chosen to be the local Fermi momentum related to the isoscalar density through
\begin{equation}
   k \equiv k_F(\bm{R}) = \biggl(\frac{3\pi^2}{2}\rho_0(\bm{R})\biggr)^{1/3} \;,
   \label{LDA}
\end{equation}
although other choices are possible that include additional $\tau$- and $\Delta\rho$-dependencies~\cite{campi77}.  The functions $\{{\cal P}_n
(\bm{R}), {\cal Q}_m (\bm{R})\}$ denote various local densities and their gradients $\{\rho_t(\bm{R}), \tau_{t}(\bm{R}), \bm{J}_{t}(\bm{R}), \bm{\nabla} \rho_t (\bm{R}), \Delta \rho_t(\bm{R})\}$, which for time-reversal invariant systems are defined by
\begin{eqnarray}
 \rho_{t}(\bm{R})  &\equiv& \rho_t (\bm{r}_1, \bm{r}_2) |_{\bm{r}_1=\bm{r}_2=\bm{R}} \;, \\
 \tau_{t}(\bm{R}) &\equiv&  \bm{\nabla}_1 \bm{\cdot} \bm{\nabla}_2 \,
  \rho_t (\bm{r}_1, \bm{r}_2) |_{\bm{r}_1=\bm{r}_2=\bm{R}} \;,\\
 \bm{J}_{t}(\bm{R}) &\equiv& - \frac{i}{2} (\bm{\nabla}_1 - \bm{\nabla}_2)\times  \bm{S}_{t} (\bm{r}_1,\bm{r}_2) |_{\bm{r}_1=\bm{r}_2=\bm{R}} \;,
\end{eqnarray}
where the isospin index $t = \{0,1\}$ labels isoscalar and isovector densities, respectively. For example, the isoscalar local density is the sum $\rho_0 =\rho_n+\rho_p$ of neutron $\rho_{n}$ and proton $\rho_{p}$ densities, while the isovector local density is the difference $\rho_1 =\rho_n-\rho_p$. Analogous expressions hold for  $\tau_t$, $\bm{J}_t$, and all other quantities labelled by the index $t =\{0,1\}$.

Applying the expansion in Eqs.~\eqref{approxscalar}--\eqref{approxvector} to the non-local exchange (Fock) energy gives a spatial integral over a sum of bilinear  and trilinear products of local densities.  For time-reversal invariant systems (truncating the expansion to second-order in gradients), the NN exchange energy becomes
\begin{eqnarray}
\label{eq:E2N}
  E_{x}^{{\rm NN}}[\rho] &\approx& \sum_{t=0,1}\int d\Rvec \biggl\{ g_t^{\rho\rho}\rho_t^2
   + g_t^{\rho\tau}\rho_t\tau_t + g_t^{\rho\Delta\rho}\rho_t\Delta\rho_t
   \nonumber \\
&&\qquad\qquad\quad \null + g_t^{J\nabla\rho}\bm{J}_t\!\bm{\cdot}\!\bm{\nabla}\rho_t
   + g_t^{JJ}\bm{J}_t^2\biggr\}\,,
\label{eq:EDF}
\end{eqnarray}
while the NNN contribution yields
\begin{eqnarray}
\label{eq:E3N}
E^{{\rm 3N}}_{x}[\rho] &\approx& \int \!d\Rvec \biggl \{ g^{\rho_0^3}\rho_0^3
  + g^{\rho_0\rho_1^2}\rho_0\rho_1^2 +  g^{\rho_0^2\tau_0}\rho_0^2\tau_0
  \nonumber \\
  & & \null +  g^{\rho_1^2\tau_0}\rho_1^2\tau_0
    +  g^{\rho_0\rho_1\tau_1}\rho_0\rho_1\tau_1
    +  g^{\rho_0^2\Delta\rho_0}\rho_0^2\Delta\rho_0
   \nonumber \\
  & & \null +   g^{\rho_1^2\Delta\rho_0}\rho_1^2\Delta\rho_0
       +  g^{\rho_0\rho_1\Delta\rho_1}\rho_0\rho_1\Delta\rho_1
   \nonumber \\
  & & \null + g^{\rho_0J_0^2}\rho_0\bm{J}_0^2
      + \, g^{\rho_0J_1^2}\rho_0\bm{J}_1^2
    + g^{\rho_1J_0J_1}\rho_1\bm{J}_0\bm{\cdot}\bm{J}_1
   \nonumber \\
  & & \null  + g^{\rho_0\bm{\nabla}\rho_0J_0}\rho_0\bm{\nabla}\rho_0\bm{\cdot}\bm{J}_0\
       \nonumber \\
  & & \null +  g^{\rho_0\bm{\nabla}\rho_1J_1}\rho_0\bm{\nabla}\rho_1\bm{\cdot}\bm{J}_1
     +  g^{\rho_1\bm{\nabla}\rho_0J_1}\rho_1\bm{\nabla}\rho_0\bm{\cdot}\bm{J}_1
     \nonumber \\
  & & \null +  \, g^{\rho_1\bm{\nabla}\rho_1J_0}\rho_1\bm{\nabla}\rho_1\bm{\cdot}\bm{J}_0
    +  g^{\rho_0^2\bm{\nabla} J_0}\rho_0^2\bm{\nabla}\bm{\cdot}\bm{J}_0
     \nonumber\\
  & & \null + \, g^{\rho_1^2\bm{\nabla} J_0}\rho_1^2\bm{\nabla}\bm{\cdot}\bm{J}_0
     + g^{\rho_0\rho_1\bm{\nabla} J_1}\rho_0\rho_1\bm{\nabla}\bm{\cdot}\bm{J}_1 \biggr \} \;,
\end{eqnarray}
where, for simplicity, the $\Rvec$-dependence of the local densities and DME couplings has been omitted.
The $\Rvec$-dependence (or equivalently, isoscalar density-dependence via Eq.~\eqref{LDA}) of the couplings arises from
the integral of the finite-range NN and NNN interactions over various products of the $\Pi$-functions, e.g.,
\begin{equation}
  g_t^{\rho\tau}(\Rvec) \sim \int dr\,r^2\,\Pi_0^{\rho}(k_Fr)
    \,\Pi_2^{\rho}(k_Fr)\, \Gamma_c^{xt}(r)  \;,
\end{equation}
where in this example $\Gamma_c^{xt}(r)$ is the central component of the exchange force $V(r)P_{\sigma}P_{\tau}$.

If the objective is to derive a fully microscopic and quantitative EDF free from any fitting to data, then the purely non-empirical HF functional of Ref.~\cite{Gebremariam:2010ni} is inadequate since un-evolved chiral interactions generate too-strong coupling between low and high momenta for HF to be a reasonable zeroth-order approximation. Moreover, it is known that even if the interactions are softened by evolving to low momentum, it is still necessary to go to at least 2nd-order MBPT to obtain a reasonable description of bulk properties of infinite matter as well as binding energies and charge radii of closed-shell nuclei.

Unfortunately, a consistent extension of the DME procedure beyond the Hartree-Fock level of MBPT has not yet been formulated. At this point in time, any attempt to microscopically construct a {\it quantitative} Skyrme-like EDF must therefore inevitably resort to either some {\it ad hoc} approximations (e.g., neglecting state-dependent energy denominators) when applying the DME to iterated contributions beyond the HF level, and/or to the re-introduction of some phenomenological parameters to be adjusted to data~\cite{negele72,negele75,hofmann97,Kaiser:2002jz,Kaiser:2009me}. 
An example of the latter approach has been recently proposed in Refs.~\cite{Gebremariam:2009ff,Gebremariam:2010ni}.

\subsection{Semi-phenomenological DME functional}

Schematically, the EFT NN and NNN potentials have the following structure
\beq
  V_{{\rm EFT}} = V_{\pi} + V_{{\rm ct}} \;,
\eeq
where $V_{\pi}$ denotes finite-range pion-exchange interactions and $V_{{\rm ct}}$ denotes scale-dependent zero-range contact terms encoding the effects of integrated-out degrees of freedom (e.g., heavier meson exchanges, high-momentum two-nucleon states, etc.) on low-energy physics.  Consequently, each DME coupling in Eqs.~\eqref{eq:E2N}--\eqref{eq:E3N} decomposes into a density-independent coupling constant arising from the zero-range contact interactions, and a density-dependent coupling function arising from long-range pion exchanges, e.g.,
\beq
g^{\rho\tau}_t  \equiv  g^{\rho\tau}_t(\Rvec;V_{\pi}) + C^{\rho\tau}_t(V_{{\rm ct}}) \;,
\eeq
and so on.
Note that the zero-range $V_{{\rm ct}}$ generates Hartree-Fock contributions that are identical in form to the standard Skyrme functional, which is hardly surprising since such functionals were originally derived as the Hartree-Fock energy density resulting from a zero-range Skyrme ``force" or pseudo-potential.

Based on the clean separation between long- and short-distance physics at the HF level, a semi-phenomenological approach where the long-distance couplings ($g_t^{m}(\Rvec;V_{\pi})$) are kept as is, and the zero-range $C^{m}_t$ are optimized to finite nuclei and infinite nuclear matter properties was suggested in Refs.~\cite{Gebremariam:2009ff,Gebremariam:2010ni}. While this is an admittedly empirical procedure, it is motivated by the observation that the dominant bulk correlations in nuclei and nuclear matter are primarily short-ranged in nature, as evidenced by Brueckner-Hartree-Fock (BHF) calculations where the Brueckner G-matrix ``heals'' to the free-space interaction at sufficiently large distances.

Therefore, while a Hartree-Fock calculation using un-evolved chiral NN and NNN interactions would provide a very poor description of nuclei, the application of the DME to such contributions nevertheless captures at least some of the non-trivial density dependencies that would arise from the finite-range tail of an in-medium vertex (e.g., a G-matrix or a perturbative approximation thereof) that sums ladder diagrams in a more sophisticated many-body treatment. In this sense, one can loosely interpret the refit of the Skyrme constants to data as approximating the short-distance part of the G-matrix with a zero-range expansion through second order in gradients. In the following section, we describe how free parameters entering the volume part of the proposed functional can be fixed to equilibrium properties of infinite matter, while the remaining free surface parameters can be fixed to properties of nuclei using a restricted SVD optimization procedure.

\section{Implementation of the DME functional} \label{sec2}

\subsection{Notations} \label{sec2sub1}

We write the proposed semi-phenomenological DME functional
\begin{equation}
  E[\rho]=\int {\cal H}(\bm{r})\, d \bm{r} \;,
 \label{EUED}
\end{equation}
in the following form
\begin{eqnarray}
  {\cal H}(\bm{r}) &=&  \frac{\hbar^2}{2m}\tau_0+ \sum_{t t'}{\cal H}_{t t'}(\bm{r}) \;,
  \\
  {\cal H}_{tt'}(\bm{r}) &=&  U_{tt'}^{\rho^2} \rho_t \rho_{t'}
   + U_{tt'}^{\rho\tau} \rho_t \tau_{t'} +  U_{tt'}^{J^2} \bm{J}_t \bm{\cdot} \bm{J}_{t'}
  \nonumber \\
  & & \null + U_{tt'}^{\rho\Delta\rho} \rho_t\Delta\rho_{t'}
      +  U_{tt'}^{\rho\nabla J}  \rho_t\bm{\nabla} \bm{\cdot} \bm{J}_{t'} \;,
\label{UED}
\end{eqnarray}
where the notation reflects that our attention is restricted to the ground states of even-even nuclei in the present paper.  Consequently, only terms built out of time-even densities are shown explicitly. The density-dependence of the couplings has been omitted for brevity.
Note that the strange ``off-diagonal'' isospin structure in Eq.~\eqref{UED} is a consequence of absorbing an extra factor of $\rho_0$ or $\rho_1$ into the definition of the $U^{m}_{tt'}$ couplings, which allows the trilinear 3N contributions in Eq.~\eqref{eq:E3N} to be written in terms of the more familiar bilinear products of local densities, e.g.,
\begin{equation}
  g^{\rho_1^2\tau_0}\rho_1^2\tau_0 = \bigl\{g^{\rho_1^2\tau_0}\rho_1\bigr\}\rho_1\tau_0
    \equiv U^{\rho\tau}_{10}\rho_1\tau_0
    \;,
\end{equation}
and so on.

The $U^{m}_{tt'}$ couplings (where $m$ runs over the bilinears $\{\rho_t\rho_{t'}, \rho_t\tau_{t'}, {\bm J}_t\bm{\cdot}{\bm J}_{t'},\rho_t\Delta\rho_{t'}, \rho_t\bm{\nabla\cdot J}_{t'}\}$) have the following general structure
\begin{eqnarray}
  U_{tt'}^{m} &=& \bigl(C_{t}^{m}+ g_{t}^{m}(u) + \rho_0~ h_{t}^{m}(u)\bigr) \delta_{t,t'}
   \nonumber \\
  & & \null + \rho_1~ h_{tt'}^{m}(u) \left(1-\delta_{t,t'}\right) \;,
  \label{urr}
\end{eqnarray}
where $u \equiv k_F(\Rvec)/\mpi$. The functions $g_{t}^{m}(u)$ are obtained by applying the DME to the Fock-energy contributions from the finite-range pion-exchange parts of the chiral EFT NN interaction through N$^2$LO.  The functions $h_{tt'}^{m}(u)$ and $h_{t}^{m}(u)$ originate from the finite-range part of the leading chiral NNN interaction (which appears at N$^2$LO), and are related to the couplings in Eq.~\eqref{eq:E3N} by
\bea
h_t^{m} &=& \rho_0\, g^{\rho_0m} ,
    \quad m\,\in\,\{\rho^2_t, \rho_t\tau_t,\rho_t\Delta\rho_t,\ldots\} \;, \\
h_{tt'}^{m}&=& \rho_1\, g^{\rho_1m} ,
    \quad m\,\in\,\{\rho_t\rho_{t'}, \rho_t\tau_{t'},\rho_t\Delta\rho_{t'}, \ldots\} \;.
\eea
The $C_{t}^{m}$ parameters correspond to the zero-range $V_{{\rm ct}}$ contributions, which as discussed in the previous section will be released for optimization. In this way, the proposed DME functional splits into two terms,
\begin{equation}
 E[\rho]=E_{\rm ct}[\rho]+E_{\pi}[\rho] \;,
  \label{dme2}
\end{equation}
where the first term $\displaystyle E_{\rm ct}[\rho]=E[\rho]_{g=h=0}$ collects all contributions from the contact part of the interaction plus higher-order short-range contributions encoded through the optimization to nuclei and nuclear matter, while the second term $\displaystyle  E_{\pi}[\rho]=E[\rho]_{C^n_{tt'}=0}$ collects  the long-range NN and NNN pion exchange contributions at the Hartree-Fock level.

This leads to the following explicit form of the DME-based energy density
\begin{equation}
  {\cal H}(\bm{r}) = \frac{\hbar^2}{2m}\tau_0+{\cal H}_{0}(\bm{r})
    +{\cal H}_{1}(\bm{r})+{\cal H}_{2}(\bm{r}) \; ,
\label{dme1}
\end{equation}
where, for $t=\{0,1\}$,
\begin{eqnarray}
  {\cal H}_{t}(\bm{r}) &=& \bigl(C_{t0}^{\rho^2}+C_{tD}^{\rho^2} \rho_0^{\gamma}
   + g_t^{\rho^2}(u)+\rho_0 h_t^{\rho^2}(u)\bigr)\rho_t^2 \nonumber \\
  & & \null + \bigl(C_t^{\rho\tau}+g_t^{\rho \tau}(u)
    +\rho_0 h_t^{\rho\tau}(u)\bigr) \rho_t\tau_t  \nonumber \\
  & & \null + \bigl(C_t^{\rho\Delta\rho}+g_t^{\rho\Delta\rho}(u)
    +\rho_0 h_t^{\rho\Delta\rho}(u)\bigr)\rho_t\Delta\rho_t \nonumber \\
  & & \null +\bigl(C_t^{\rho\nabla J}+g_t^{\rho\nabla J}(u)
    +\rho_0 h_t^{\rho\nabla J}(u)\bigr)\rho_t \nabla J_t  \nonumber \\
  & & \null + \bigl(C_t^{J^2}+g_t^{J^2}(u)+\rho_0 h_t^{J^2}(u)\bigr) J_t^2
\label{urr1}
\end{eqnarray}
and
\begin{eqnarray}
  {\cal H}_{2}(\bm{r}) &=& \rho_1 h_{10}^{\rho\tau}(u)\rho_1 \tau_0
    + \rho_1 h_{10}^{\rho\Delta\rho}(u)\rho_1\Delta \rho _0 \nonumber \\
  & & \null + \rho_1 h_{10}^{J^2}(u) J_1 J_0
  + \rho_1 h_{10}^{\rho \nabla J}(u)\rho_1\nabla J_0 \;. \label{urr-cross1}
\end{eqnarray}
The explicit forms of the functions $g_{t}^{m}(u)$ and $h_{tt'}^{m}(u)$ have been given in Ref.\cite{Gebremariam:2010ni} and the companion Mathematica notebooks.  In order to gain a feeling about the new density dependencies entering through such couplings, we provide stripped-down ``skeleton expressions" along with several explicit examples in Appendix~\ref{app0}.

In order to facilitate the use of the DME functional in nuclear EDF calculations, we have also developed a general module written in FORTRAN~90~\cite{Kortelainen:2010aa},  which can easily be ported to any existing EDF solver. It contains all of the lengthy  expressions for the DME couplings  $U^m_{tt'}$, Eq.~(\ref{urr}), their functional derivatives with respect to the density matrix, and numerically stable approximate expressions at small $u$. The module also has the capability to calculate related infinite nuclear matter properties.

\subsection{Contact part}
\label{sec2sub3}

The contact part $E_{\rm ct}[\rho]$ has the form of the standard Skyrme functional
\begin{equation}
{\cal H}_{\rm ct}(\bm{r}) = \frac{\hbar^2}{2m}\tau_0+{\cal H}^{\rm ct}_{0}(\bm{r})
  +{\cal H}^{\rm ct}_{1}(\bm{r}) \;,
  \label{cp}
\end{equation}
where
\begin{eqnarray}
  {\cal H}^{\rm ct}_{t}(\bm{r}) &=& \bigl(C_{t0}^{\rho^2}
   +C_{tD}^{\rho^2} \rho_0^{\gamma}\bigr)\rho_t^2 + C_t^{\rho\tau} \rho_t\tau_t
  + C_t^{\rho\Delta\rho} \rho_t\Delta\rho_t
  \nonumber \\
  & & \null +C_t^{\rho\nabla J} \rho_t \nabla J_t
    + C_t^{J^2} J_t^2 \;.
\label{cpt}
\end{eqnarray}
This is illustrated in Appendix~\ref{app1},  where the link between the coupling constants and the historical $(\text{t}_{n},\text{x}_{n})$-parameterization of Skyrme ``forces" is explicitly given.

As for the standard Skyrme functional, Eq.~\eqref{cp} contains 13 parameters,
\begin{equation}
  \{C_{t0}^{\rho^2}, C_{{tD}}^{\rho^2}, C_t^{\rho\Delta\rho}, C_t^{\rho\tau},C_t^{J^2},
    C_t^{\rho\nabla J},   \gamma\} \;,
\label{c-parameters}
\end{equation}
which are to be released for optimization to infinite matter and finite nuclei properties.  While these parameters have exactly the same form as in the standard Skyrme functional, the existence of the long-range part in the functional will obviously modify their optimized values.

\subsection{The parameter $\gamma$}
\label{gamma}

Early versions of the Skyrme functional motivated the $ \rho^{\gamma}$ term appearing in Eq.~\eqref{cpt} as arising from a zero-range NNN force, in which case $\gamma \equiv 1$. However, this interpretation was soon found to be problematic, as $\gamma=1$ yields too large an incompressibility~\cite{bender03b}. Subsequent Skyrme parameterizations largely cured this difficulty by letting $\gamma$ float, with values typically between 1/6 and 1/3.

In EFT studies of dilute Fermi systems interacting with zero-range interactions, one finds similar non-integer powers of $\rho$ appearing in the energy density, which can be traced to correlation (i.e., beyond HF) effects~\cite{Hammer:2000xg}. Even in this much simpler model system, where a controlled and well-defined EFT expansion is possible, it is interesting to note that one finds {\it multiple} non-integer powers of $\rho$ occurring at low orders in the expansion. Given that the nuclear many-body problem is much more complicated, with many additional possible sources of non-analytic behavior due to the interplay of finite-range NN and NNN interactions and short-range correlation effects analogous to those found in the dilute fermion system, the single non-integer $\rho^{\gamma}$ term in Eq.~\eqref{cp} is probably not justified on microscopic grounds.

Nevertheless, in the short term we follow standard practice with a single $\rho^{\gamma}$ term in the functional, which in our case can be loosely viewed as parameterizing the HF contribution of the NNN contact term, plus higher-order correlation effects that are implicitly included in the refit to data. However, ultimately  one would like to revisit this issue to see if MBPT can be used to provide insight regarding the form of such non-analytic terms.

\subsection{Finite-range part}
\label{sec2sub4}

The finite range part $E^{\pi}[\rho]$ follows from
\begin{equation}
  {\cal H}_{\pi}(\bm{r}) = {\cal H}^{\pi}_{0}(\bm{r})
     + {\cal H}^{\pi}_{1}(\bm{r})+{\cal H}^{\pi}_{2}(\bm{r}) \;,
  \label{dme2b}
\end{equation}
where, for $t=\{0,1\}$,
\begin{eqnarray}
{\cal H}^{\pi}_{t}(\bm{r}) &=& \bigl(g_t^{\rho^2}(u)+\rho_0 h_t^{\rho^2}(u)\bigr)\rho_t^2 \nonumber \\
 & & \null + \bigl(g_t^{\rho \tau}(u)+\rho_0 h_t^{\rho\tau}(u)\bigr) \rho_t\tau_t\nonumber \\
 & & \null + \bigl(g_t^{\rho\Delta\rho}(u)+\rho_0 h_t^{\rho\Delta\rho}(u)\bigr)\rho_t\Delta\rho_t \nonumber \\
 & & \null + \bigl(g_t^{J^2}(u)+\rho_0 h_t^{J^2}(u)\bigr) J_t^2 \nonumber \\
 & & \null + \bigl(g_t^{\rho\nabla J}(u)+\rho_0 h_t^{\rho\nabla J}(u)\bigr)\rho_t \nabla J_t
 \;,
\label{urr2}
\end{eqnarray}
and
\begin{eqnarray}
{\cal H}^{\pi}_{2}(\bm{r}) &=& \rho_1 h_{10}^{\rho \tau}(u)\rho_1 \tau_0
   + \rho_1 h_{10}^{\rho\Delta\rho}(u)\rho_1\Delta \rho _0 \nonumber \\
 & & \null + \rho_1 h_{10}^{J^2}(u) J_1 J_0 +
   \rho_1 h_{10}^{\rho \nabla J}(u)\rho_1\nabla J_0 \;.
   \label{urr-cross2}
\end{eqnarray}
Couplings entering $E^{\pi}[\rho]$ are entirely determined in terms of the finite-range NN and NNN interaction parameters, and are therefore frozen during the optimization procedure. In the present work, the values for the couplings that enter the finite-range chiral EFT interactions are taken from Ref.~\cite{Epelbaum:2004fk}.

\subsection{Hartree NN contributions}
\label{sec2sub5}

In general, it is possible to apply the DME to both Hartree and Fock energies so that the complete Hartree-Fock energy is mapped into a local functional. It is known since the original work of Negele and Vautherin, however, that treating the Hartree contributions exactly provides a better reproduction of the density fluctuations and the energy produced from an exact HF calculation \cite{negele75,sprung75}. Restricting the DME to the exchange contribution significantly reduces the self-consistent propagation of errors~\cite{negele75}. Moreover, treating the Hartree contribution exactly generates no additional complexity in the numerical solutions of the resulting self-consistent equations compared to applying the DME to both Hartree and Fock terms. Lending further support to Negele and Vautherin's conclusions, we find that the present DME-based functional becomes extremely susceptible to numerical instabilities when the DME approximation is used for the Hartree terms, which immediately disappear when the Hartree terms are treated exactly.

In the present work, a simplification is used such that finite-range NN Hartree contributions are treated in the Local Density Approximation (LDA). An exact treatment of these contributions and their influence on the results will be examined in a future investigation of the DME-based functional.

\section{Constraining the volume term parameters } \label{sec3}
\subsection {INM with the DME-based functional}

We turn now to a discussion of how equilibrium properties of infinite nuclear matter (INM) can be used to fix the 7 free volume parameters $\{C^{\rho^2}_t, C^{\rho^2}_{tD}, C_t^{\rho\tau}, \gamma\}$ in the functional. In INM, the total energy per particle defines the saturation curve $W(\rho_n,\rho_p)$.   Its derivation discards the Coulomb energy and all gradient terms that are zero for a homogeneous system, and substitutes the kinetic energy density with its Thomas-Fermi expression, which is exact in this case. Assuming spin saturation, one also disregards the terms involving the spin-orbit current density, ${\bm J}_t$.

The expansion of $W(\rho_n,\rho_p)$ around the equilibrium density $\rho_c$ in a Taylor series in  $\rho=\rho_n+\rho_p$ and $I \equiv (\rho_n-\rho_p)/\rho$ yields
\begin{eqnarray}
 W(\rho,I)&=&  W(\rho)+S_2(\rho) I^2+ S_4(\rho) I^4 \;, \label{INMexpansion}
  \\
  W(\rho) &=&  \frac{E^{\rm NM}}{A} + \frac{P^{\rm NM}}{ \rho_c^2 } \delta \rho
   + \frac{K^{\rm NM}}{18 \rho_c^2} \left( \delta \rho \right)^2  \;,
  \\
  S_2(\rho) &=&  a_{\rm sym}^{\rm NM}+\frac{L^{\rm NM}}{3 \rho_c}\delta \rho
  + \frac{\Delta K^{\rm NM}}{18 \rho_c^2} \left( \delta \rho \right)^2 \;,
\label{NMT}
\end{eqnarray}
where $\delta \rho=\left( \rho-\rho_c \right)$, while ${E^{\rm NM}}/{A}$,  $P^{\rm NM}$, $K^{\rm NM}$,  $a_{\rm sym}^{\rm NM}$,  $L^{\rm NM}$ and $\Delta K^{\rm NM}$ denote the total energy per nucleon at equilibrium, the pressure, the nuclear matter incompressibility, the symmetry energy coefficient, the density derivative of $a_{\rm sym}^{\rm NM}$, and the isovector correction to the incompressibility at saturation density $\rho_c$ of nuclear matter, respectively. The quartic term in $I$ can be safely neglected in Eq.~\eqref{INMexpansion} in practice.

The INM equation of state (EOS) following from the functional Eq.~\eqref{UED} takes the form
\begin{eqnarray}
 W(I,\rho) &=&  \frac{\hbar^2}{2m} \tau_0
 \nonumber \\
 & & \null + \bigl(C_{00}^{\rho^2}+C_{0D}^{\rho^2} \,
   \rho^{\gamma}+g_0^{\rho^2}(u)+\rho \, h_0^{\rho^2}(u)\bigr)\rho
  \nonumber \\
 & & \null + \bigl(C_{10}^{\rho^2}+C_{1D}^{\rho^2} \,
   \rho^{\gamma}+g_1^{\rho^2}(u)+\rho \, h_1^{\rho^2}(u)\bigr) I^2 \rho
   \nonumber \\
 & & \null + \bigl(C_0^{\rho\tau}+g_0^{\rho\tau}(u)+\rho \, h_0^{\rho\tau}(u)+I^2 \rho \,
    h_{10}^{\rho\tau}(u)  \bigr) \tau_0
   \nonumber \\
 & & \null + \bigl(C_1^{\rho\tau}+g_1^{\rho\tau}(u)+\rho \,
      h_1^{\rho\tau}(u)\bigr) I \, \tau_1 \;,
\label{wCa}
\end{eqnarray}
where
\begin{eqnarray}
  u  &=& \displaystyle  \frac{k_F}{m_\pi}=\frac{1}{m_\pi}\left(\frac{3 \pi^2}{2}\right)^{1/3} \rho^{1/3} \;,
  \\
\tau_0 &=&  \displaystyle  \frac{1}{2} C \rho^{2/3}
   \left((1 + I)^{5/3} + (1 - I)^{5/3}\right) \;,  \label{eq:tau0}
  \\
\tau_1 &=&  \displaystyle  \frac{1}{2} C \rho^{2/3}
   \left((1 + I)^{5/3} - (1 - I)^{5/3}\right) \;,
  \\
   C &=& \displaystyle \frac{3}{5} \left(\frac{3 \pi^2}{2}\right)^{2/3} \;.
\label{wF}
\end{eqnarray}
Our strategy is to  express unknown volume parameters in terms of nuclear matter equilibrium quantities $\rho_c$, ${E^{\rm NM}}/{A}$, $K^{\rm NM}$, $a_{\rm sym}^{\rm NM}$, $L^{\rm NM}$, $m_s^{*}$, and $m_v^{*}$, where $m_s^{*}$ and $m_v^{*}$ are the isoscalar and isovector effective masses, respectively. This strategy has been recently applied in the context of the optimization of pure Skyrme functionals \cite{Kortelainen:2010hv}.

\subsection{Symmetric nuclear matter constraints} \label{sec3sub1}

Symmetric nuclear matter (SNM) is characterized by equal neutron and proton densities $\rho_n=\rho_p=\rho/2$, where $\rho$ is the isoscalar density. All isovector terms are thus zero. The isoscalar kinetic energy density per particle from Eq.~\eqref{eq:tau0} is
\begin{equation}
  \tau= C \rho^{2/3} \;.
  \label{tauTF}
\end{equation}
Parameters $C_{00}^{\rho^2}$, $C_{0D}^{\rho^2}$ and  $C_0^{\rho\tau}$ are then expressed in terms of
${E^{\rm NM}}/{A}$,  $\rho_c$, and $m_s^{*}$ through
\begin{eqnarray}
 C^{\rho^2}_{00} &=& \frac{1}{3\gamma\rho_c}  \bigl[ 3 (\gamma +1) \frac{E^{\rm NM}}{A}-\frac{\hbar^2}{2m} \left(3- \left(2-3 \gamma\right)  m^{*\, -1}_s\right)\tau _c
   \nonumber \\ & & \qquad\qquad \null + A_{00}(u_c) \bigr]
     \; , \label{CfromSNMa} \\
   C^{\rho^2}_{0D}&=& \frac{1}{3 \gamma \rho _c^{\gamma +1} } \bigl[-3 \frac{E^{\rm NM}}{A}-\frac{\hbar^2}{2m} \left(2m^{*\, -1}_s-3 \right) \tau _c
    \nonumber \\ & & \qquad\qquad \null + A_{0D}(u_c) \bigr]
     \;, \\
C^{\rho \tau}_0&=&  \frac{\hbar^2}{2m} \left(m^{*\, -1}_s-1\right)\frac{1}{\rho _c}-g_0^{\rho \tau }(u_c)-h_0^{\rho \tau}(u_c) \rho _c  \;,
\label{CfromSNM}
\end{eqnarray}
where  $\tau_c$ and $u_c$ are the kinetic energy density  and the dimensionless Fermi momentum at the saturation density $\rho_c$, respectively. Explicit expressions for $A_{t0}(u)$  and $A_{tD}(u)$ are given in Appendix~\ref{app2}.

As for the parameter  $\gamma$, one can either leave it as a free parameter or eliminate it using the incompressibility $K^{\rm NM}$. The resulting expression  is
\begin{equation}
\gamma = \frac{
   -K^{\text{\rm NM}}
   -9 E^{\rm NM}/A+\frac{\hbar^2}{2m}  (4 m^{*\, -1}_s-3 )\tau _c+ A_{ \gamma }(u_c)
   }{
   9 E^{\rm NM}/A+3 \frac{\hbar^2}{2m}  (2 m^{*\, -1}_s-3 )\tau _c+B_{ \gamma }(u_c)
   }
   \;, \label{eq:gamma}
\end{equation}
where explicit expressions for $A_{ \gamma }(u_c)$ and $B_{ \gamma }(u_c)$ are given in Appendix~\ref{app2}.

\subsection{Asymmetric nuclear matter constraints} \label{sec3sub2}

In the regime of isospin asymmetric INM, neutron and proton densities are different  ($\rho_0 =\rho, \rho_1 = I \rho$) and the isovector terms contribute. One can therefore express parameters $C_{10}^{\rho^2}, C_{1D}^{\rho^2}$ and $C_1^{\rho\tau}$ in terms of   $a_{\rm sym}^{\rm NM}$, $L^{\rm NM}$ and $m_{v}^{*}$ through
\begin{eqnarray}
  C^{\rho^2}_{10} &=&
  \frac{
        27 (\gamma +1) a_{\rm sym}^{\rm NM} -9 L^{\rm NM}+20
        (2-3 \gamma)   C^{\rho \tau}_0  \rho _c \tau _c
   }{27 \gamma  \rho _c} \nonumber \\
  & & \null +
  \frac{
        \frac{\hbar^2}{2m} \bigl( (9 \gamma-6) m^{*\, -1}_v -12\gamma +5\bigr) \tau _c
        +A_{10}(u)
   }{27 \gamma  \rho _c} \;, \nonumber \\
   & &  \\
  C^{\rho^2}_{1D} &=&
     \frac{-27 a_{\rm sym}^{\rm NM}+9 L^{\rm NM}
     -40 C^{\rho \tau}_0 \rho _c \tau _c
    }{27 \gamma   \rho _c^{\gamma +1} } \nonumber \\
        & & \null +
     \frac{\frac{\hbar^2}{2m} (30 m^{*\, -1}_v-25) \tau _c
      + A_{1D}(u)
    }{27 \gamma   \rho _c^{\gamma +1} } \;, \\
   C^{\rho \tau}_1  &=&   \frac{\hbar^2}{2m}(m^{*\, -1}_s-m^{*\, -1}_v)\frac{1}{\rho _c}
   -g^{\rho \tau}_1(u)- h^{\rho \tau}_1(u)\rho _c \;,
\label{CfromANM}
\end{eqnarray}
where $C_0^{\rho\tau}$ has already been  determined by Eqs.~(\ref{CfromSNM}).

\subsection{Reference SLy4 properties}
\label{sec4sub2}

\begin{figure}[tbh-]
\centering
{\includegraphics[scale=0.55]{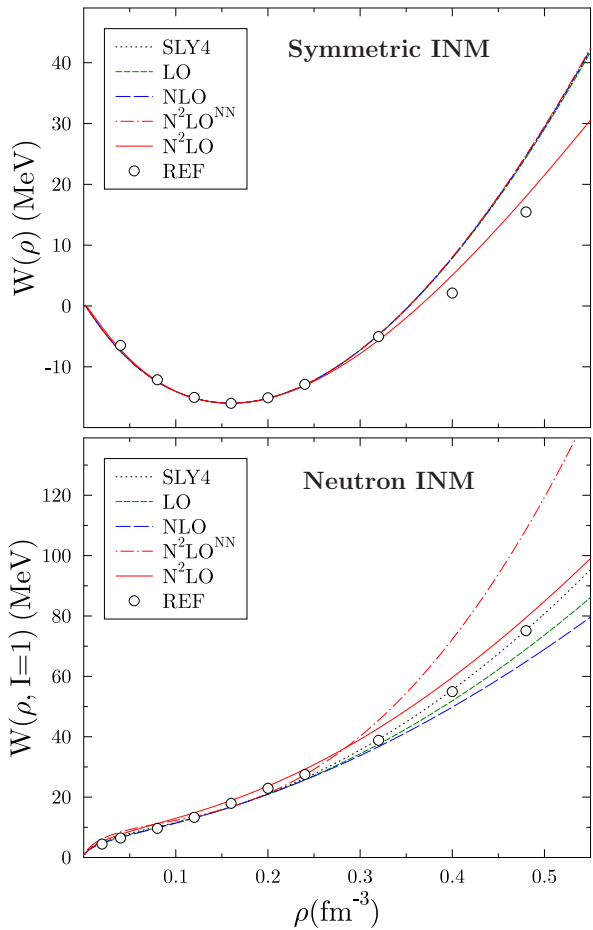}}
\caption{(Color online) INM saturation curves  for symmetric (top panel) and neutron (bottom panel) nuclear matter calculated with the standard SLy4 Skyrme functional  (dotted lines), and the DME  functional at LO (dashed  lines), NLO (dot-dashed lines), N$^2$LO without NNN contributions (short dashed lines), and N$^2$LO (solid lines). Ab-initio results~\cite{Akmal:1998cf} are plotted as reference points. All functionals are constrained to reproduce the reference INM saturation values from Eq.~\eqref{NMSLy4}.}
\label{fig2}
\end{figure}

In this work, we will often use as benchmark the SLy4 parametrization of the Skyrme force \cite{(Cha98)}. The optimization protocol of this interaction included data obtained from ab initio calculations in nuclear matter of Ref.~\cite{Akmal:1998cf}, and we see in Fig.~\ref{fig2} that the saturation curves obtained with SLy4 agree well with the ab initio results. We therefore take the INM equilibrium characteristics of the SLy4 parameterization as an acceptable set of values to be used in Eqs.~\eqref{CfromSNMa}--\eqref{CfromANM},
\begin{eqnarray}
  \frac{E^{\rm NM}}{A} &=&  -15.97\, \text{MeV}, \quad\
    \rho_c = 0.1595\,\text{fm}^{-3}, \nonumber \\
 K^{\rm NM} &=&  229.9\,\text{MeV}, \qquad
    m_{s/v}^{*\, -1} = 1.44/1.25 \;, \label{NMSLy4} \\
  a_{\rm sym}^{\rm NM}  &=& 32.0\,\text{MeV} , \qquad\
     L^{\rm NM} =  45.96 \,\text{MeV} .  \nonumber
\end{eqnarray}
The resulting parameters $C_{t0}^{\rho^2}$, $C_{t{\rm D}}^{\rho^2}$, $C_t^{\rho\tau}$ and $\gamma$ are compared with the original SLy4 parameters in Table~\ref{table1}, and the associated INM curves are shown in Fig.~\ref{fig2} for symmetric matter (top panel) and pure neutron matter (bottom panel).

As seen from Table~\ref{table1}, the values of the refitted parameters differ substantially from the original SLy4 values. The original value of $\gamma=1/6$, for example, increases to about $\gamma=1/3$ when only NN contributions are taken into account, but becomes almost equal to one when both NN and NNN contributions are accounted for at the N$^2$LO level. Nevertheless, the EOS for the DME-based functional are practically identical to the original SLy4 curves for symmetric (top panel of Fig.~\ref{fig2}) and pure neutron matter (bottom panel of Fig.~\ref{fig2}) for densities relevant to nuclei.

At higher densities ($\rho > 0.3$ fm$^{-3}$) the symmetric matter EOS (Fig.~\ref{fig2} top) remains completely predetermined by the imposed equilibrium values  Eq.~\eqref{NMSLy4} with a slight deviation towards the reference points when NNN contributions are taken into account (N$^2$LO).  Deviations in this density range become more visible for the neutron matter EOS (Fig.~\ref{fig2} bottom), as isovector properties depend on the EFT order and/or whether the NNN force is taken into account (N$^2$LO) or not  (N$^2$LO$^{\rm NN}$).

\begin{table}[tbh-]
 \caption{\label{table1} Parameters entering the volume part of the standard Skyrme functional and the DME functional at LO, NLO, N$^2$LO without NNN contributions (N$^2$LO$^{\rm NN}$), and N$^2$LO, all reproducing the same values Eq.~(\ref{NMSLy4}) for INM saturation properties.}
\begin{ruledtabular}
\begin{tabular}{lrrrrr}
Parameters &SLy4 & LO  &  NLO  &  N$^2$LO$^{\rm NN}$   &N$^2$LO \\
\hline
$C^{\rho^2}_{00}$  &-933.342   & -727.093  &-757.689& -777.805&-607.108 \\
$C^{\rho^2}_{10}$  & 830.052   &  477.931  &477.931& 677.296&331.438 \\
$C^{\rho^2}_{0D}$  & 861.062   &  612.114  &628.504& 641.017&-1082.85 \\
$C^{\rho^2}_{1D}$  & -1064.273 & -705.739  &-694.665&-952.381&-4383.27 \\
$C^{\rho \tau}_0$     & 57.129    &  33.885   &18.471& 26.0411&322.4 \\
$C^{\rho \tau}_1$     & 24.657    &  32.405   &92.233& -51.8352&-156.90 \\
$\gamma$              &  0.16667  &  0.30622 &0.287419& 0.275049&1.06429  \\
\end{tabular}
\end{ruledtabular}
\end{table}

If one does not impose  the nuclear incompressibility value $K^{\rm NM}=229.9~\text{MeV}$  but instead varies the value of $\gamma$, one can trace the influence of the DME contributions on the INM incompressibility $K^{\rm NM}$. Such results are shown in Table~\ref{table2} for two values of $\gamma$:  $\gamma=1/6$, corresponding to the SLy4 parameterization,  and
$\gamma=1$, originally proposed for Skyrme functionals~\cite{vautherin72a}.

\begin{table}[tbh-]
\caption{\label{table2} Nuclear matter incompressibility $K^{\rm NM}$ (in $\text{MeV}$) calculated at two different values of $\gamma$.}
\begin{ruledtabular}
\begin{tabular}{lccccc}
&SLy4& $LO$&NLO&N$^2$LO$^{\rm NN}$ &N$^2$LO \\
\hline
$\gamma=1/6$ &229.90& 208.49& 211.42& 213.34 & 440.50 \\
$\gamma=1$ &356.36& 336.34& 338.96& 340.68  & 244.98 \\
\end{tabular}
\end{ruledtabular}
\end{table}

Contributions from the NN interaction generally reduce the value of $K^{\rm NM}$ by about 10--20\,MeV. Conversely, NNN contributions give too-high values for $K^{\rm NM}$  unless  $\gamma\sim 1$, which brings the N$^2$LO $K^{\rm NM}$ value to the physically accepted range of 220--250\,MeV. Interestingly, the fact that the preferred value of $\gamma$ is rather close to 1  when finite-range NNN contributions are explicitly accounted for seems to contradict the naive argument recalled in Sec.~\ref{gamma} that $\gamma$ should encode both the Hartree-Fock contribution of the NNN contact term ($\gamma=1$) but also higher-order correlation effects producing non-integer values for $\gamma$.

\subsection{First test on surface parameters}
\label{sec5sub1}

The equilibrium INM properties allow us to constrain the zero-range volume parameters of the DME-based functional (Table~\ref{table1}), but not the parameters $C^{\rho \Delta \rho}_{t}$ and $C^{\rho \nabla J}_{t}$ entering the surface part of the functional or the tensor parameters $C^{J^2}_t$ as their associated terms are zero in spin symmetric nuclear matter. We note right away that all tensor terms from both pion exchanges and the Skyrme-like contact terms are omitted in the present proof-of-principle investigation. The reasons for that are briefly discussed in Sec.~\ref{sec6sub1}. 

As for a first test on the surface parameters of the DME-based functional, we keep the volume ones at the values estimated from INM (Table~\ref{table1}), and simply set $C^{\rho \Delta \rho}_{t}$ and $C^{\rho \nabla J}_{t}$  equal to zero. That is, we let the long-range DME part of the functional ($E_{\pi}[\rho]$) generate all of the surface contributions.  Results of such calculations at LO  for two benchmark nuclei  $^{40}$Ca and $^{208}$Pb are shown in the second column of Table~\ref{table3}. The comparison with the results by SLy4 (the first column in Table~\ref{table3}) shows unacceptable over-binding of about 140  MeV and 320 MeV in  $^{40}$Ca  and $^{208}$Pb, respectively.

\begin{table}[htb]
 \caption{\label{table3} Comparison of calculated SLy4 energies (in MeV) with the results from DME calculations at LO for nuclei $^{40}$Ca and $^{208}$Pb: kinetic energy for neutrons $T_{n}$ and protons $T_{p}$, volume energy $E_{V}$, surface energy $E_{S}$ (see also Table \ref{table4}) and total energy $E_{T}$.. Volume DME parameters are taken from Table~\ref{table2}. Surface DME parameters $C_t^{\rho\Delta\rho}$ and $C_t^{\rho \nabla J}$  are set equal to zero (third column) or to their SLy4 values (fourth column). }
\begin{ruledtabular}
\begin{tabular}{lrrr}
\noalign{\smallskip}
      \multicolumn{2}{c}{}   & \multicolumn{1}{c}{$C_t^{\rho\Delta\rho}$ and $C_t^{\rho \nabla J}=0$} &    \multicolumn{1}{c}{\text{ from SLy4}}     \\
\noalign{\smallskip}
\hline
  &SLy4~~& DME:LO & DME:LO \\
\hline
\noalign{\smallskip}
      \multicolumn{4}{c}{$^{40}$Ca}      \\
\noalign{\smallskip}
$ T_n   $  &     321.788     &  401.334     &  313.497        \\
$ T_p   $  &     313.215     &  393.286     &  304.782       \\
$ E_{V} $ &   -1161.116   &  -1362.847  &  -1130.494   \\
$ E_{S}$  &     111.046     &   7.889         &  109.992        \\
$ E_{T} $ &    -344.227     &  -480.770    &  -332.150    \\ \hline
\noalign{\smallskip}
      \multicolumn{4}{c}{$^{208}$Pb}      \\
\noalign{\smallskip}
$ T_n   $  &     2527.799  &  2819.082   &  2490.574    \\
$ T_p   $  &     1336.341  &  1481.145   &  1321.254   \\
$ E_{V} $ &    -6514.517 &  -7092.892  &  -6429.071  \\
$ E_{S}$  &      315.116   &   65.083        &   316.181    \\
$ E_{T} $ &    -1635.160  & -1950.690    & -1599.798  \\
\end{tabular}
\end{ruledtabular}
\end{table}

As a second test case, we compute the LO DME results in which we have taken the SLy4 values for  $C^{\rho \Delta \rho}_{t}$ and $C^{\rho \nabla J}_{t}$ (third column on Table~\ref{table3}). In this case, the results are much closer but now an under-binding is seen of about 10  MeV and 30 MeV in $^{40}$Ca  and $^{208}$Pb, respectively.

Table~\ref{table3} suggests that it should be possible to optimize the surface parameters in the DME functional in a manner similar to the 
optimization done for standard Skyrme functionals. Broadly speaking, one could think of procedures based on semi-infinite NM properties, or on the leptodermous expansion of the functional. Both approaches would fix the parameters entering the surface part of the functional on a set of well-defined surface coefficients. Alternatively, one can probe and constrain the surface parameters using properties of finite nuclei. This is the path taken in the present work and described in Section~\ref{sec6}.

\subsection{Stability of the DME-based functional}
\label{sec5sub2}

Our preliminary analysis of calculations using different sets of $C^{m}_t$ parameters has shown that the DME functional is somewhat more sensitive to instabilities than standard Skyrme functionals, especially when the N$^2$LO NNN contributions are taken into account. Nevertheless, we have been able to sidestep these issues thus far with minimal modifications.

The DME procedure, for example, contains a freedom in the choice of the coordinate system when expanding the one-body density matrix with respect to the center-of-mass coordinate~\cite{Koehl95}. This freedom introduces a factor $(a^2-a+1/2)$ in front of the surface terms proportional to $\rho \Delta \rho$, with $a$ ranging between zero and one. In quantum chemistry studies of molecular exchange energies, taking $a=0$, which corresponds to expanding $\rho(\rvec_1,\rvec_2)$ asymmetrically about $\rvec_2$, was found to be the optimal choice by a large margin~\cite{Koehl95}. For nuclei, we find that the quantum chemistry choice leads to overly strong surface contributions, resulting in heavy under-binding, which can only be compensated by substantially reducing the associated surface constants $C_t^{\rho \Delta \rho}$. Such a reduction, however, leads to numerical instabilities of the HFB solution for practically all  nuclei studied.

In our test calculations, we are able to avoid such difficulties by using the original Negele-Vautherin choice ($a = 1/2$), which corresponds to symmetrically expanding $\rho(\rvec_1,\rvec_2)$ about the center-of-mass coordinate $\Rvec = (\rvec_1+\rvec_2)/2$.  This choice minimizes the long-range $\rho \Delta \rho$ contributions, and leads to stable results for the nuclei studied in the present paper.

\begin{figure}[htb]
\centering
{\includegraphics[scale=0.55]{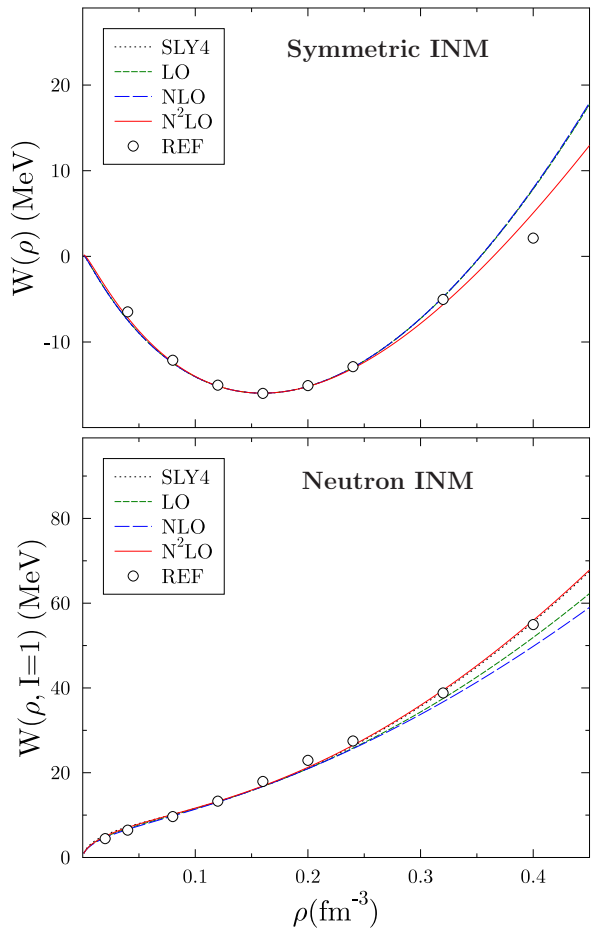}}
\caption{(Color online) INM saturation curves  for symmetric  (top panel) and neutron (bottom panel) nuclear matter  calculated with the modified SLy4$'$ Skyrme functional  (dotted lines) and the  DME  functional at LO (dashed lines), NLO (dot-dashed lines), N$^2$LO (solid lines) using the parameters from Table~\ref{table4}.}
\label{fig4}
\end{figure}

Another source of instabilities has been found with respect to the isovector behavior of the DME-based functional. For example, increasing the value of the symmetry energy parameter $a_{\rm sym}^{\rm NM}$ generates a functional which has an instability that cannot be compensated by the contact part of the functional. As a matter of fact, the N$^2$LO functional displays systematic instabilities in finite nuclei calculations when it is defined with the values of Table~\ref{table2}. However, by slightly modifying the values of INM characteristics to $a_{\rm sym}^{\rm NM}=30$~ MeV and  $L^{\rm NM}=40$~ MeV (compare with values in Eq.~(\ref{NMSLy4})), the N$^2$LO functional becomes stable enough to carry out the SVD optimization of the surface parameters as discussed in the next section. The resulting EOS in both symmetric and pure neutron matter are shown on Fig.~\ref{fig4} where one sees that the latter is much better behaved at high density as one increases the EFT order.

In the calculations presented below, we have managed thus far to avoid such instabilities in our zeroth-order optimization (``pre-optimization'') to fix the volume and surface parameters of the DME-based functional.  However, a more sophisticated global optimization of the DME-based functional will ultimately require a more precise and detailed analysis of its stability properties to rule out the most common problems~\cite{Kor10}.

\section{Pre-optimization} \label{sec6}

A global optimization of a given EDF parameterization becomes a very involved procedure as soon as one extends the data set to include observables beyond nuclear binding energies. Such optimization procedures are expensive, as they require a large number of functional evaluations. It is always useful for the global optimization to have preliminary estimates for the values of the functional parameters.

In the present section, we describe a particularly convenient method to generate approximate estimates for the surface parameters constrained to binding energies and odd-even mass (OEM) differences. The method uses the explicit linear dependence of the functional on the parameters  $C^m_t$. The procedure can be seen as a particular implementation of the optimization algorithm based on the regression analysis of Refs.~\cite{bertsch05a,Toivanen:2008im}.

\subsection{SVD optimization procedure} \label{sec6sub1}

We first define the $\chi^{2}$ function to be minimized. Due to the limitations of the optimization method, see below, the experimental data set is reduced to nuclear binding energies (i.e.\ masses) and pairing gaps. Except for proton radii, the data set is the same as in Ref.~\cite{Kortelainen:2010hv}. The $\chi^{2}$ function is then defined as
\begin{equation}
  \chi^2 = \frac{1}{n_{d}-n_{x}}\sum_{i=1}^{n_{d}}
\left( \frac{E^{\rm exp}_{i}-E_{i} }{ \sigma_{i}} \right)^2 \;,
\label{EAchi}
\end{equation}
where $n_{d}$ denotes the total number of experimental data points, $n_{x}$ the number of free parameters,
$E^{\rm exp}_{i}$ the experimental values while $E_{i}$ are the corresponding calculated ones.
The weights $\sigma_{i}$ are 2~MeV for binding energies and 50~keV for pairing gaps. The same weights were used in Ref.~\cite{Kortelainen:2010hv}.

The binding energy $E_{A}$ computed from a Skyrme-like EDF depends linearly on the coupling constants such that one can write
\begin{equation}
  E_{A}=e^0_{A}+\sum_{k=1}^{N} C_k e^k_{A} \;,
\label{EA}
\end{equation}
where the index $k \mapsto \{m,t\}$ runs over all $N$ combinations of
$m=\{\rho^2, \rho\tau, \rho\Delta\rho, \rho \nabla J,  J^2 \}$ and isospin $t=0,1$.
Furthermore,  $e^0_{A}$ collects all contributions to the energy that do not explicitly depend on the
functional parameters  $\{C_k\}$, i.e., kinetic and Coulomb energies as well as contributions coming from
finite-range pion exchanges. The terms $e^k_{A}$ entering Eq.~(\ref{EA}) are integrals multiplying the associated parameter $C_k$ in
the EDF. For example, the term associated with $C^{\rho\tau}_0$ is
\begin{equation}
\displaystyle e^{\rho_{0}\tau_{0}}_{A} = \int\! \rho_0(\bm{r}) \tau_0(\bm{r})\, d\bm{r} \;,
\label{ek0}
\end{equation}
where the densities are computed for nucleus $A$. Strictly speaking, $e^k_{A}$ depends on the value of the coupling constants. However, in close vicinity of the local minimum the behavior of the functional can be approximated to be linear in the couplings.

Starting with a given initial set of coupling constants $\{ C^{(0)}_k \}$, one can perform EDF calculations for all the nuclei belonging to the data set. This provides values of the integrals $e^k_{A}$ for every $k$ and $A$. If some of the coupling constants are
left out from the optimization process, i.e.\ $n_{x}<N$, their contribution to the total energy can be transferred to the constant
$e^0_{A}$. Substituting coefficients $e^k_{A}$ in Eq.~(\ref{EA}) and requesting the minimum of the $\chi^{2}$ which now contains linearized
energies, one ends up with a system of $n_{d}$ linear equations for the $n_{x}$ unknown parameters $\{ C_k \}$:
\begin{equation}
  \sum_{k=1}^{n_{x}} C_k e^k_{A_i}=E^{\rm exp}_{A_i}-e^0_{A_i},~(i=1,\ldots,n_{d} > n_{x}) \;.
\label{EAsol}
\end{equation}
Solving the system of equations (\ref{EAsol}) with respect to $\{ C_k \}$ using the SVD method gives the solution for the minimum of the $\chi^2$ \cite{[Pre92]} under the linear hypothesis. Due to the previously mentioned (small) non-linearities in $e^k_{A}$, the process has to be repeated iteratively until the true local minimum is found.

The procedure described above can be easily extended to functionals containing a pairing contribution. In the present work we include
a mixed delta pairing~\cite{(Dob02)},
\begin{equation}
  E^{pp}=\frac{1}{2}\int  \left(1-\frac{\rho_{0}}{2\rho_{c}}\right)  \left(V_n \tilde{\rho}_n^2+V_p \tilde{\rho}_p^2\right)  d\bm{r} \;,
\label{pair}
\end{equation}
where $\rho_{c}=0.16$ fm$^{-3}$, and where $V_q$, $\tilde{\rho}_q$ denote the pairing strength and pairing density for neutrons ($q=n$)
and protons ($q=p$), respectively. In EDF calculations with pairing, one can calculate average pairing gaps for neutrons ($\Delta_n$) and protons
($\Delta_p$) using
\begin{equation}
  \Delta_q=\frac{V_q}{N_q} \int  \left(1-\frac{\rho (\bm{r})}{2\rho_0}\right)  \rho_q(\bm{r}) \tilde{\rho}_q(\bm{r}) \;,
\label{dpair}
\end{equation}
where $N_n$, and $N_p$ are the number of neutrons and protons, respectively.

It is straightforward to add the pairing parameters $V_n$ and $V_p$ to the SVD optimization procedure above because Eqs.~(\ref{pair}) and (\ref{dpair}) are linear with respect to them. The pairing energy Eq.~(\ref{pair}) should be added to Eq.~(\ref{EA}), while Eq.~(\ref{EAsol}) should be extended with the requirement that the calculated (neutron or proton) pairing gaps Eq.~(\ref{dpair}) for selected nuclei should approximate the odd-even mass (OEM) difference defining the experimental pairing gap $\Delta_q^{A_j {\rm exp}}$
\begin{equation}
  \Delta_q^{A_j {\rm exp}}=\Delta_q^{A_j}\;, \quad (j=1,...,M_q) \;.
\label{DqAsol}
\end{equation}

Besides the adjustment to binding energies and OEM differences, it is necessary to impose boundaries to the domain of variation of the free parameters due to the approximate nature of the nuclear functional and the incomplete set of experimental observables used.
For example, one cannot release for optimization parameters $C^{\rho \nabla J}_t$ controlling spin-orbit contributions simultaneously with parameters $C^{J^2}_t$ governing tensor contributions~\cite{Lesinski:2007zz}. In the present proof-of-principle investigation, we drop all tensor contributions from pion exchanges entering the DME-based functional.
That is, the amplitudes $U_{tt'}^{JJ}$ are set to zero so that we can avoid optimizing the corresponding contact terms. Eventually, such terms will be subject to detailed investigation for the reasons explained in the Introduction. Even then, all remaining parameters $C^{m}_t$ cannot be fully released for a SVD optimization. Such attempts lead to a negative isovector effective mass (experimental data cannot constrain this quantity \cite{Kortelainen:2010hv}) or to a too-small INM saturation density (charge radii are not included in the present optimization).

\subsection{Preliminary parameterization}
\label{sec6sub3}

The SVD optimization procedure has been performed with respect to the six ($N=6$) parameters $C^{\rho \Delta \rho}_{0}$, $C^{\rho \Delta \rho}_{1}$, $C^{\rho \nabla J}_{0}$, $C^{\rho \nabla J}_{1}$, $V_n$, and $V_p$ using the binding energies of 30 spherical and 42 deformed nuclei (i.e., $M=72$), neutron pairing gaps of 4 nuclei ($M_n=4$), and proton pairing gaps of another 4 nuclei ($M_p=4$). Nuclear properties have been calculated as in Ref.~\cite{Kortelainen:2010hv} using the HFBTHO solver~\cite{stoitsov05a} in the Lipkin-Nogami regime of approximate particle number projection~\cite{stoitsov07a}. For comparison, the same SVD optimization procedure has also been performed for the standard Skyrme functional starting from the SLy4 parameterization. The results define a new parameterization referred to as SLy4$'$.

\begin{table}[bth]
 \caption{\label{table4} Parameters entering the volume, surface, and pairing parts of the standard Skyrme functional and the DME-based functional at LO, NLO, and N$^2$LO. The associated $\chi^2$ values  and root-mean-square deviations  (RMSD) are also shown.}
\begin{ruledtabular}
\begin{tabular}{lrrrrr}
Parameter &SLy4~~& SLy4$'$~~&  LO  &   NLO  &  N$^2$LO  \\
\hline \noalign{\smallskip}
 & \multicolumn{5}{c}{Volume Parameters}      \\
$C^{\rho^2}_{00}$  & \multicolumn{2}{r}{ -933.342~~~~~~~~}    &  -727.093   &-757.689    &-607.108 \\
$C^{\rho^2}_{10}$  & \multicolumn{2}{r}{ 830.052~~~~~~~~}     &  474.871    &477.931       &316.939 \\
$C^{\rho^2}_{0D}$  &\multicolumn{2}{r}{ 861.062~~~~~~~~}      &  612.104    &628.504      &-1082.854 \\
$C^{\rho^2}_{1D}$  & \multicolumn{2}{r}{ -1064.273~~~~~~~~}  & -705.739   &-694.665    &-4369.425 \\
$C^{\rho \tau}_0$    &\multicolumn{2}{r}{ 57.129~~~~~~~~}         &  33.885     &18.471        &322.4 \\
$C^{\rho \tau}_1$    &\multicolumn{2}{r}{ 24.657~~~~~~~~}        &  32.405      &92.233        &-156.901 \\
$\gamma$                &\multicolumn{2}{r}{ 0.16667~~~~~~~~}      &  0.30622   &0.287419    &1.06429  \\
\noalign{\smallskip} \hline \noalign{\smallskip}
 & \multicolumn{5}{c}{Surface Parameters}      \\
$C^{\rho \Delta \rho}_{0}$  &-76.287   &-76.180                              & -67.437      &-63.996     &-197.132   \\
$C^{\rho \Delta \rho}_{1}$  &15.951     &24.823                               & 21.551       &-9.276       &-12.503     \\
$C^{\rho \nabla J}_{0}$      &-92.250    &-92.959                             &-95.451       &-95.463     &-193.188 \\
$C^{\rho \nabla J}_{1}$     &-30.75       &-82.356                             &-65.906       &-60.800      &37.790   \\
\noalign{\smallskip} \hline \noalign{\smallskip}
 & \multicolumn{5}{c}{Pairing Parameters}      \\
$V_n$                                   &-258.992   &-232.135                           &-241.203    &-241.484    &-272.164 \\
$V_p$                                   &-258.992   &-244.050                           &-252.818    & -252.222   & -286.965 \\
\noalign{\smallskip} \hline \noalign{\smallskip}
 & \multicolumn{5}{c}{  SVD Optimization Results}      \\
$\chi^2$                               &12.5002      &2.1235                            & 1.837          & 1.7662     &1.7884  \\
$(E)_{\rm RMSD} $                         &7.008           &2.6931                            & 2.5539        & 2.5143    &2.590    \\
$(\Delta_n)_{\rm RMSD}$              &0.1297        &0.0828                            & 0.0587        & 0.0554     &0.0476  \\
$(\Delta_p)_{\rm RMSD}$              &0.094          &0.0988                             & 0.0902       & 0.0866     &0.0706  \\
\end{tabular}
\end{ruledtabular}
\end{table}

The values of the parameters resulting from the SVD optimization, together with the $\chi^2$ values and the resulting root mean square deviations (RMSD), are shown in Table~\ref{table4}.
The first column in Table~\ref{table4} shows the results for SLy4. Since our HFBTHO calculations are performed under the Lipkin-Nogami procedure, SLy4 leads to quite a high value of $\chi^2 \approx 12.5$. The second column in Table~\ref{table4} shows the results for SLy4$'$. The resulting $\chi^2$ is about six times smaller, $\chi^2 \approx 2.12$.

The optimal DME-based functional further reduces the value of $\chi^2$, but by a much less significant amount. This is nevertheless a remarkable result keeping in mind the extremely involved structure, i.e.\ the rich density-dependence of the coupling functions, of the DME contributions and the fact that they do not contain optimization parameters.  Interestingly, the N$^2$LO parameterization (last column of Table~\ref{table4}) performs as well as standard Skyrme functionals with a reasonable incompressibility $K^{\rm NM}=230$ MeV, even though its contact part is characterized by a density-dependent power $\gamma\sim 1$.

Because the DME-based functional is found to provide as good (or slightly better) a description of bulk properties of finite nuclei as standard Skyrme functionals, a refined (and much more costly) global optimization can be undertaken that will eventually lead to a systematic analysis of its ability to improve the known deficiencies of standard functionals. Such a task is left for a future work. With a much less ambitious objective in mind, the following section gives some insight into the novelties that could be expected from DME functionals in typical nuclear structure applications by collecting a sample of results obtained with the parametrization of Table \ref{table4}.

\section{Selected results in finite nuclei}
\label{sec7}

This section summarizes results for selected nuclei, comparing their properties  calculated with the standard Skyrme functional (SLy4$'$ set of parameters) and the DME-based functional  with the parameters given in  Table~\ref{table4}. Let us recall again that a detailed comparison of the DME functional to experimental data is premature until a more rigorous global optimization is performed.

\begin{figure}[b]
\centering
\includegraphics[scale=0.5]{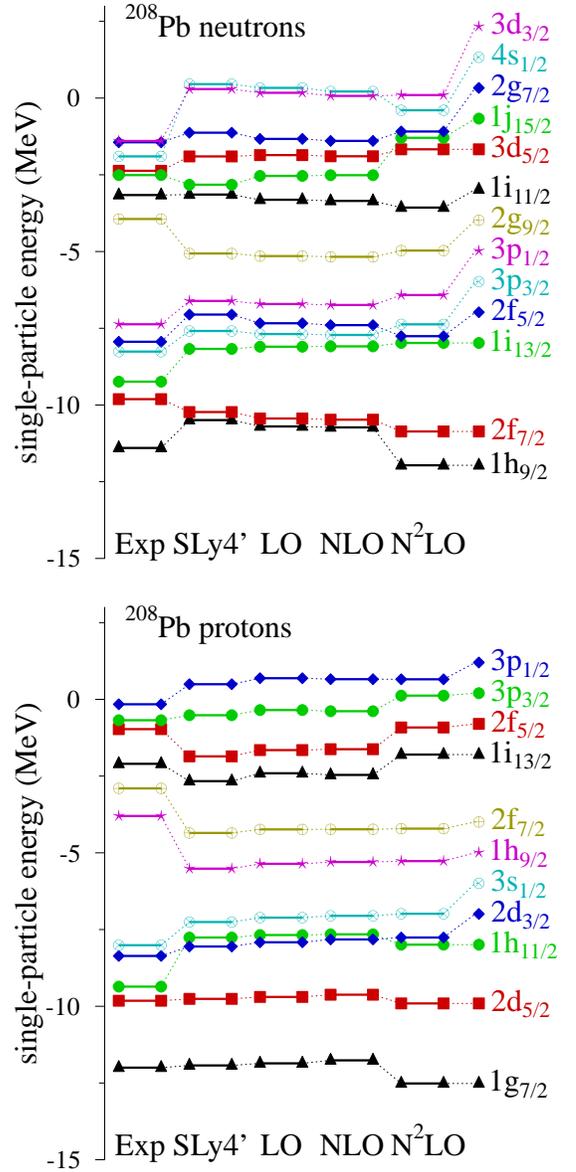} 
\caption{(Color online) Comparison of $^{208}$Pb experimental single-particle energies (in MeV) for neutrons and protons with canonical single-particle energies, calculated with the standard SLy4$'$ functional and the  LO, NLO, and N$^2$LO  DME functionals.}
\label{fig5}
\end{figure}

As the first example, Fig.~\ref{fig5} compares experimental single-particle energies of~\cite{Sch07} for the nucleus $^{208}$Pb  with the canonical single-particle energies calculated with SLy4$'$ and the  LO, NLO, and N$^2$LO  DME functionals. In general, the comparison shows that the DME functional does not modify the Skyrme results significantly. The largest deviations are seen in the N$^2$LO case mainly due to the stronger spin-orbit contact contribution (compare the values of $C^{\rho \nabla J}_{t}$ from Table~\ref{table4}). In \cite{Kor08} it was found that Skyrme functionals perform poorly on single-particle energies. The marked differences between SLy4$'$ and N$^2$LO single-particle spectra indicate that this situation may be improved.

However, as discussed in the previous section, all tensor contributions (contact term and finite-range contributions) have been set to zero to avoid the unnaturally large and strongly cancelling values  for $C^{\rho\nabla J}_t$ and $C^{J^2}_t$ that arise in the present optimization protocol. Since the  interplay between tensor and spin-orbit terms is crucial to understanding the evolution of nuclear shell structure, a detailed comparison of single-particle energies to data and standard Skyrme functionals is not appropriate with the present restricted optimization.

\begin{figure}[bht-]
\centering
\includegraphics[scale=0.5]{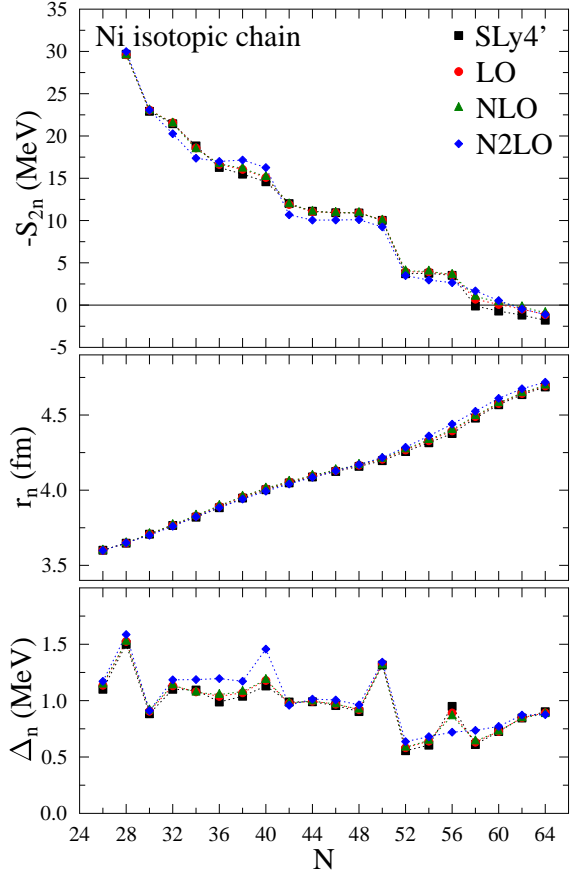}  
\caption{(Color online) Two-neutron separation energies (top panel), neutron rms radii (middle panel), and average neutron pairing gaps (bottom panel) for the Ni isotopic chain of nuclei in the region of the neutron drip line, calculated with the standard SLy4$'$ functional and the LO, NLO, and N$^2$LO  DME-based functionals. }
\label{fig6}
\end{figure}

Similar comparisons between SLy4$'$, LO, NLO, and N$^2$LO results are shown in Fig.~\ref{fig6} for the two-neutron separation energies (top panel), neutron rms radii (middle panel) and the average neutron gaps (bottom panel) for nuclei in the Ni chain in the region up to the neutron drip line. Again, the comparison in Fig.~\ref{fig6} shows similar behaviors for the Skyrme and DME-based functionals, with the largest deviations coming at N$^2$LO. Let us remember, though, that small differences in separation energies can play an important role for reliable predictions of the position of the neutron drip line. As in the case of different Skyrme parameterizations, the DME functionals could lead to a shift in this prediction.

The fact that the Skyrme and DME functionals produce very similar results for this pool of observables is in fact encouraging, since we do not want to lose the good features that phenomenological functionals based on the Skyrme force have acquired over the years. On the other hand, it would also be disappointing if the rich, microscopically-derived and non-trivial density dependence of the DME functionals could not bring in new physics that can not be captured by Skyrme functionals.  

\begin{figure}[htb]
\centering
\includegraphics[scale=0.4]{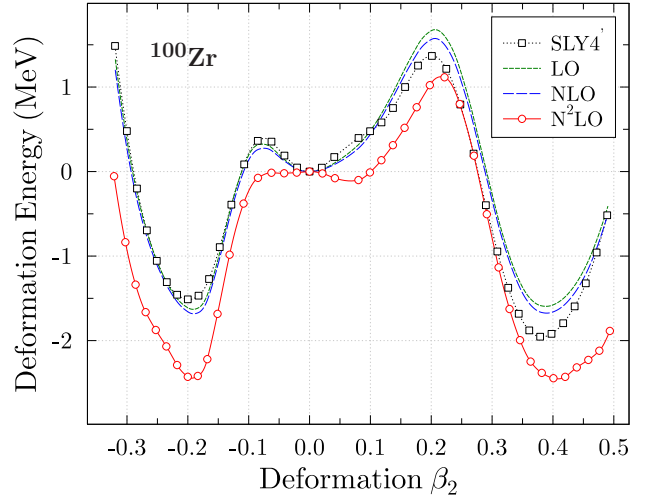}  
\caption{(Color online) Deformation energy for the nucleus $^{100}$Zr calculated with standard SLy4$'$ functional (dotted line, squares) and
the DME functional in LO (dashed line), NLO (dot-dashed line), and N$^2$LO  (solid line, circles). }
\label{fig7}
\end{figure}

In this respect, nuclear deformation is an excellent probe, as it reflects the competition between the bulk properties of the interaction and its single-particle content. Located right after the onset of deformation, the nucleus $^{100}$Zr is characterized by the coexistence of 3 minima, oblate, spherical and prolate, the relative position of which is highly sensitive to the interaction. Fig.~\ref{fig7} shows that, in contrast to the Skyrme functional, the oblate and prolate minimum for DME functionals have almost the same energy (shape coexistence). At N$^2$LO, the difference is even more marked, as the spherical minimum disappears and is shifted at small prolate deformation. This behavior of the DME functionals can probably be related to a combination of small differences in single-particle energies of closed-shell nuclei, see Fig. \ref{fig5}, as well as rather different surface bulk properties, see the value of coupling constants in Table \ref{table4}. Indeed, it is a particularity of that new generation of EDF parameterizations to provide surface and spin-orbit terms with density-dependent couplings.

\begin{figure}[htb]
\centering
\includegraphics[scale=0.4]{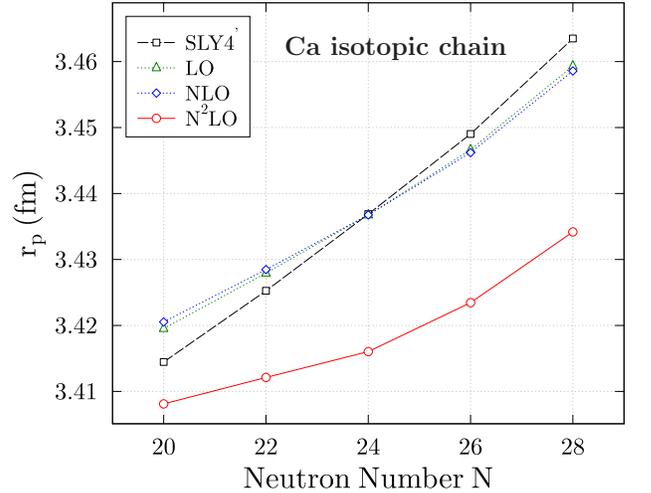}  
\caption{(Color online) Comparison of the proton rms radii for nuclei along the Ca-chain calculated with the standard SLy4$'$ functional (squares) and the DME functional in LO (dashed line), NLO (dot-dashed line), and N$^2$LO  (solid circles). }
\label{fig8}
\end{figure}

Another example of systematic differences is seen in the proton rms radii along the Ca isotopic chain as shown on Fig.~\ref{fig8}. Since r.m.s. radii were not included in the pre-optimization of the DME functional, the particular value of the proton radius is irrelevant. However, the isotopic trend is a marker for the iso-vector channels of the functional, and the differences of slopes between the Skyrme and DME functionals, and curvature between LO, NLO and N$^{2}$LO might be indicative of new physics. 

Whether these changes improve or deteriorate the quality of the current functional with respect to the experimental data is irrelevant, as the parametrization of Table \ref{table4} should only be thought of as a prototype. A detailed study of the capabilities of DME functionals to reproduce experimental data, as well as more systematic comparison with standard parametrizations of the Skyrme functional, will be performed once a more comprehensive optimization procedure, such as the one used in Ref.~\cite{Kortelainen:2010hv}, has been carried out.

\section{Conclusions}
\label{sec8}

In the present paper, we have given a practitioner's view of how the microscopically motivated DME functional of Gebremariam {\it et al.}~\cite{Gebremariam:2010ni}, which possesses a much richer set of density dependencies than traditional Skyrme functionals,  can be implemented in existing Skyrme HFB codes. Empirical infinite nuclear matter properties are used to constrain the volume parameters of the Skyrme-like part, followed by a restricted singular value decomposition (SVD) optimization procedure to fix its density-independent surface parameters.  We find that the proposed functional gives numerically stable results and exhibits a small but systematic reduction in $\chi^2$ compared to standard Skyrme functionals, thus justifying its suitability for future global optimizations and large-scale calculations.

The DME-based functional takes the same general form as standard Skyrme functionals, with the key difference that each coupling is composed of a density-independent Skyrme-like piece that is optimized to data, plus a density-dependent coupling function determined solely (no free parameters) from the HF contributions of the underlying finite-range NN and NNN interactions.  In this way, the functional is split into a parameter-free ``long-range'' piece that is directly linked to underlying NN and NNN pion-exchange contributions (treated at the HF level), and a short-range piece that is identical in form to the standard Skyrme functional with parameters that are optimized to data.

After reviewing the structure of the DME-based functional and motivating the semi-phenomenological approach used in the present work, it was demonstrated how the free contact parameters entering the volume part of the functional can be eliminated in a one-to-one fashion in terms of equilibrium infinite nuclear matter characteristics. The influence of the finite-range DME contributions to symmetric and neutron INM was investigated with the most significant modification seen in the N$^2$LO case when the three-body interaction is taken into account. In this case, we find a reasonable incompressibility $K^{\rm NM}=230$~MeV with a Skyrme parameter
$\gamma\sim 1$, a result which cannot be achieved within the standard Skyrme functional.

A preliminary (pre-) optimization procedure for surface and pairing parameters using the SVD-optimization algorithm was performed using the binding energies of 72 spherical and deformed nuclei,  as well as 8 odd-even mass differences.  It  was found that the pre-optimized DME-based functional performs as well or slightly better than the standard Skyrme functional with respect to the optimized binding energies and OEM differences. The same is true also for other nuclear characteristics, e.g., nuclear rms radii, pairing gaps, separation energies, single-particle energies, as well as for nuclei that are not included in the optimization. These preliminary results are very encouraging, as they imply that more elaborate global optimizations of the DME functional will, at the very worst, preserve the already impressive level of performance provided by traditional Skyrme functionals, and very likely improve on them.

The present results have been obtained under two restrictions, which are not expected to modify our general conclusions and which will be removed in a future work. First, the NN Hartree contributions have been treated within the local density approximation. However, they can be easily taken into account exactly, as the computational cost is the same as for the calculation of the Coulomb direct term, which is already included in nuclear EDF calculations.

Much more important, insofar as it plays a central role in future investigations of the spectroscopic and single-particle properties of the new functional, is the neglect of the tensor contributions in the present study. The issue is that in the present optimization procedure, the contact tensor coupling constant cannot be released for optimization together with the spin-orbit coupling constant, as the optimization drives both couplings to unnaturally large (and strongly canceling) values. It should be noted that this difficulty is not specific to the DME-based functional, as similar issues arise with the standard Skyrme functional.  The next step, removing both limitations mentioned above, is to apply a complete optimization procedure with the DME-based functional and perform systematic comparisons to the standard Skyrme functional and experimental data. Work in this direction is already in progress.

\section*{Acknowledgements}

We thank W. Nazarewicz, T. Papenbrock and T. Lesinski for useful discussions. This work was supported by the Office of Nuclear Physics, U.S. Department
of Energy under Contract Nos. DE-FC02-09ER41583 (UNEDF SciDAC Collaboration),
DE-FG02-96ER40963, DE-FC02-07ER41457, DE-FG02-07ER41529 (University of Tennessee),  DE-FG0587ER40361 (Joint Institute
for Heavy Ion Research), and DE-FC02-09ER41585 (Michigan State University) and the National Science Foundation
under Grant Nos.~PHY--0653312 and PHY--0758125. Computational resources were provided through an INCITE award ``Computational
Nuclear Structure" by the National Center for Computational Sciences (NCCS) and
the National Institute for Computational Sciences (NICS) at Oak Ridge National Laboratory.

\appendix

\section{ \label{app0} DME skeleton expressions}
The lengthy analytic expressions for the DME couplings tend to obscure their underlying structural simplicity. Therefore, it is more illuminating to display the couplings in a ``skeleton form''
that still conveys its non-trivial density dependence.

The DME coupling $g_t^{m}(u)$ is given as a sum of LO, NLO, and N$^2$LO
contributions (recall $u = k_F/m_\pi$),
\begin{equation}
  g_t^{m}(u)=\left. g_t^{m}(u) \right| _{\rm LO}+\left. g_t^{m}(u) \right| _{\rm NLO}+\left. g_t^{m}(u) \right| _{\rm N^2LO},
  \label{gsmall1}
\end{equation}
where $t=0,1$ and the index $m$ runs over the standard bilinear forms $\{\rho_t^2, \rho_t\tau_t,\rho_t\Delta\rho_t, {\bm J}_t^2,{\bm J}_t\nabla\rho_t\}$. These contributions
are of the following generic form
\begin{eqnarray}
\left. g(u) \right| _{\rm LO} &=& \alpha^g_0 + \beta^g_0 \log \bigl( 1+4 u^2\bigr)  + \gamma^g_0 \arctan(2u) \;,
  \nonumber \\  & & \\
\left. g(u) \right| _{\rm NLO}&=&\alpha^g_1+ \beta^g_1\Bigl( \log \bigl( 1+2 u^2+2u \sqrt{1+u^2}\,\bigr)\Bigr)^2
  \nonumber \\
&&  \hspace*{-.4in} \null +\gamma^g_1 \sqrt{1+u^2}  \log \bigl( 1+2 u^2+2u \sqrt{1+u^2}\,\bigr) \;,
  \\
\left. g(u) \right| _{\rm N^2LO}&=& \alpha^g_2 + \beta^g_2 \log \bigl( 1+ u^2\bigr) + \gamma_2  \arctan(u) \;,
\label{gsmall2}
\end{eqnarray}
where $\alpha^g_k=\alpha^g_k(u)$, $\beta^g_k=\beta^g_k(u)$, and $\gamma^g_k=\gamma^g_k(u)$  are rational polynomials in $u$, with  their dependence on $t$ and $m$ not explicitly shown. The explicit expressions for $k=0,1,2$ are given in Ref~\cite{Gebremariam:2010ni} and the companion Mathematica notebooks.

In a similar way, the DME couplings $h^m_{t t'}(u)$ that collect the N$^2$LO NNN contributions read
\begin{eqnarray}
  h^m_{t t'}(u)&=& \alpha^h_0 + \beta^h_0 \log \bigl( 1+4 u^2\bigr) + \beta^h_1 \bigl(\log \bigl( 1+4 u^2\bigr) \bigr)^2
   \nonumber \\
  && \null + \gamma^h_0 \arctan(u) +  \gamma^h_1 \left( \arctan(2u)\right)^2
   \nonumber \\
  && \null + \gamma^h_2  \log \bigl( 1+4 u^2\bigr)   \arctan(2u)  \;,
\label{gsmall3}
\end{eqnarray}
where the explicit expressions for the rational polynomials $\alpha^h_k=\alpha^h_k(u)$, $\beta^h_k=\beta^h_k(u)$, and $\gamma^h_k=\gamma^h_k(u)$, with  their dependence on $t$ and $m$ are not explicitly shown.

\section{ \label{app1} Coupling constants and tx-parameterization}

Using the following explicit one-to-one relation between $C^m_t$ parameters and the (t,x) Skyrme parameters,
\begin{equation}
  C_{0}^{\rho_0\rho_0} = \frac{3 \text{t}_0}{8} \;, \qquad C_{0}^{\rho_1\rho_1}=-\frac{ \text{t}_0}{4} \left(\text{x}_0+\frac{1}{2}\right) \;,
\end{equation}
\begin{equation}
C_{{\rm D}}^{\rho_0\rho_0} =  \frac{\text{t}_3}{16} \;,
  \qquad  C_{{\rm D}}^{\rho_1\rho_1}=-\frac{\text{t}_3}{24}
  \left(\text{x}_3+\frac{1}{2}\right) \;,
\end{equation}
\begin{equation}
C^{\rho_0\Delta\rho_0} = \frac{\text{t}_2}{16}  \left(\text{x}_2+\frac{5}{4}\right)-\frac{9}{64}\text{t}_1 \;,
\end{equation}
\begin{equation}
C^{\rho_1\Delta\rho_1} =  \frac{3 \text{t}_1}{32} \left(\text{x}_1+\frac{1}{2}\right)+\frac{\text{t}_2}{32}
   \left(\text{x}_2+\frac{1}{2}\right) \;,
 \end{equation}
\begin{equation}
C^{\rho_0\tau_0} =  \frac{3 \text{t}_1}{16} +\frac{\text{t}_2}{4}  \left(\text{x}_2+\frac{5}{4}\right) \;,
 \end{equation}
\begin{equation}
C^{\rho_1\tau_1} =  -\frac{\text{t}_1}{8} \left(\text{x}_1+\frac{1}{2}\right)+\frac{\text{t}_2}{8}  \left(\text{x}_2+\frac{1}{2}\right) \;,
 \end{equation}
\begin{equation}
C^{J_0^2} =  -\frac{\text{t}_1}{16}  (\text{x}_1-\frac{1}{2})-\frac{\text{t}_2}{16}  ( \text{x}_2+\frac{1}{2})+\frac{5}{32}(3~to+te) \;,
 \end{equation}
\begin{equation}
C^{J_1^2} =  \frac{1}{32}(\text{t}_1-\text{t}_2)+\frac{5}{16}(to-te) \;,
 \end{equation}
\begin{equation}
C^{\rho_0\nabla J_0} =  -b_4 - \frac{1}{2} b_4' \;,
   \qquad C^{\rho_1\nabla J_1}=- \frac{1}{2} b_4' \;,
 \end{equation}
and substituting them into Eq.~(\ref{cp}), one can verify that the contact part of the DME functional (\ref{cp}) is equivalent to the well-known Skyrme energy density:
\begin{eqnarray}
{\cal H}^c(\bm{r}) & = &  \frac{\hbar^{2}}{2m}\tau
  \nonumber \\
 & & \null +\frac{\text{t}_0}{2}  \biggl(\left(\frac{\text{x}_0}{2}+1\right)\rho^2
   -\left(\text{x}_0+\frac{1}{2}\right)\sum\limits_{q}\rho_q^2 \biggr)
  \nonumber \\
 & & \null +\frac{\text{t}_1}{4}
   \biggl(\left(\frac{\text{x}_1}{2}+1\right) \rho~ \tau -\left(\text{x}_1+\frac{1}{2}\right)\sum\limits_{q}\rho_q \tau_q \biggr)
  \nonumber \\
 & & \null +\frac{\text{t}_2}{4}
   \biggl(\left(\frac{\text{x}_2}{2}+1\right)\rho~ \tau +   \left(\text{x}_2+\frac{1}{2}\right)\sum\limits_{q}\rho_q \tau_q \biggr)
  \nonumber \\
 & & \null -\frac{3 \text{t}_1}{16}
   \biggl(\left(\frac{\text{x}_1}{2}+1\right) \rho \Delta \rho
   +\left(\text{x}_1+\frac{1}{2}\right)\sum\limits_{q}\rho_q \Delta \rho_q \biggr)
  \nonumber \\
 & & \null +\frac{\text{t}_2}{16}
   \biggl(  \left(\frac{\text{x}_2}{2}+1\right)\rho \Delta \rho
   +\left(\text{x}_2+\frac{1}{2}\right)\sum\limits_{q}\rho_q \Delta \rho_q \biggr)
  \nonumber \\
 & & \null +\frac{\text{t}_3}{12}  \biggl(\left(\frac{\text{x}_3}{2}+1\right)\rho^2 -\left(\text{x}_3+\frac{1}{2}\right) \sum\limits_{q}\rho_q^2 \biggr)\rho^{\gamma }
  \nonumber \\
 & & \null - \frac{1}{8}\bigl(\text{t}_1 \text{x}_1+\text{t}_2 \text{x}_2 -5 (\text{to}+\text{te})\bigr)  \sum\limits_{q}\bm{J}_q^2
  \nonumber \\
 & & \null -  \frac{1}{16} \left(\text{t}_1\left(\text{x}_1-1\right)
 +\text{t}_2\left(\text{x}_2+1\right)-
   10~\text{to}\right) \bm{J}_n\bm{\cdot}\bm{J}_p
  \nonumber \\
 & & \null -  (\text{b}_4 \rho~ \bm{\nabla}\bm{\cdot}\bm{J}+ \text{b}_4'\sum\limits_{q}\rho_q\bm{\nabla}\bm{\cdot}\bm{J}_q) \;,
  \label{SEDF}
\end{eqnarray}
where in neutron-proton notation $q=(n,p)$, densities without an index stand for the total densities, e.g., $\rho=\rho_n+\rho_p$, $\tau=\tau_n+\tau_p$, and $\bm{J}=\bm{J}_n+\bm{J}_p$.

\section{ \label{app2} DME functions for INM}
The explicit expression for functions appearing in the INM equations are
\begin{eqnarray}
  A_{ \gamma }(u)
   &=&  \bigl(u g_0^{' \rho }+u^2 g_0^{'' \rho }\bigr) \rho_c \nonumber \\
   & & \null + \bigl(9 h_0^{\rho }+7 u h_0^{' \rho }+u^2 h_0^{'' \rho }\bigr)\rho_c^2
   \nonumber \\
   & & \null +   \bigl(5 u g_0^{' \rho \tau }+u^2 g_0^{'' \rho \tau }\bigr)  \rho _c \tau_c
   \nonumber \\
  & & \null +  \bigl(21 h_0^{\rho \tau }+11 u h_0^{' \rho \tau }+u^2 h_0^{'' \rho \tau }\bigr) \rho_c^2 \tau_c \;,
\end{eqnarray}

\begin{eqnarray}
  B_{ \gamma }(u)&=&
      3 u  g_0^{' \rho } \rho _c +    \bigl(9 h_0^{\rho }+3 u h_0^{' \rho }\bigr) \rho _c^2
      \nonumber \\
    & & \null +       3 u  g_0^{' \rho \tau} \rho _c \tau _c
   + \bigl(9 h_0^{\rho \tau }+3 u h_0^{' \rho \tau }\bigr) \rho _c^2 \tau _c
   \;,
\end{eqnarray}

\begin{eqnarray}
  A_{00}(u) &=&  \bigl(u g_0^{' \rho }-3\gamma  g_0^{\rho }\bigr)\rho _c
   +   \bigl(u h_0^{' \rho }-3 (\gamma -1) h_0^{\rho }\bigr) \rho _c^2
     \nonumber \\
   & & \null +  u   g_0^{' \rho \tau }\rho _c \tau _c
   + \bigl(3 h_0^{\rho \tau }+u h_0^{' \rho \tau}\bigr)\rho _c^2 \tau _c
   \;,
\end{eqnarray}

\begin{eqnarray}
   A_{0D}(u) &=&  -u  g_0^{' \rho }\rho _c -\bigl(3 h_0^{\rho }+u h_0^{' \rho }\bigr)\rho _c^2  \nonumber \\
   & & \null -u  g_0^{' \rho \tau} \rho _c \tau _c-\bigl(3 h_0^{\rho \tau }+u h_0^{' \rho \tau }\bigr) \rho _c^2 \tau _c \;,
\end{eqnarray}

\begin{eqnarray}
        A_{10}(u)&=&\ 9 \left(-3 \gamma  g_1^{\rho} + u g_1^{' \rho } \right)\rho _c
	    \nonumber \\
            & & \null + 9\left( 3 (1-\gamma)  h_1^{\rho} + u h_1^{' \rho} \right)  \rho_c^2
	     \nonumber \\
             & & \null + 5\left((8-12 \gamma )  g_0^{\rho \tau }+ u (g_0^{' \rho \tau }
             +3   g_1^{' \rho \tau}) \right) \rho_c \tau_c
	     \nonumber \\
            & & \null +  \left( 5 (11-12 \gamma )  h_0^{\rho \tau }
            -27 \gamma  h_{10}^{\rho \tau} \right) \rho_c^2 \tau _c
	     \nonumber \\
            & & \null +  45 \left( h_{10}^{\rho \tau} + h_1^{\rho \tau} \right) \rho_c^2 \tau _c
	     \nonumber \\
           & & \null + u\left(5   h_0^{' \rho \tau }
           +9  h_{10}^{' \rho \tau} \right)  \rho_c^2 \tau _c
	   \nonumber \\
           & & \null +  15 u  h_1^{' \rho \tau} \rho _c^4 \tau _c^2
	   \;,
\end{eqnarray}

\begin{eqnarray}
                   A_{1D}(u)&=&\ -9 u  g^{' \rho }_1  \rho _c
                            -\ 9 \left( 3 h_1^{\rho}+u h_1^{' \rho} \right) \rho _c^2
			    \nonumber \\
                            & & \null - 5\left(
                              8 g_0^{\rho \tau }+u g_0^{' \rho \tau } +3 u g_1^{' \rho \tau}
                              \right) \rho _c \tau _c
			     \nonumber \\
                            & & \null - 5 \left(
                             11 h_0^{\rho \tau }+9h_1^{\rho \tau}+9 h_{10}^{\rho \tau}
                             \right) \rho _c^2 \tau _c
			     \nonumber \\
                            & & \null - u \left(
                             5  h_0^{' \rho \tau}+15  h^{' \rho \tau}_1+9  h_{10}^{' \rho \tau}
                             \right) \rho _c^2 \tau _c
			     \;,
\end{eqnarray}
where prime and double prime denote the first and second derivative with respect of $u$, respectively.

\end{document}